\documentclass[11pt]{article}
\usepackage[utf8]{inputenc}
\usepackage{amsfonts}
\usepackage{amsmath}
\usepackage{amssymb}
\usepackage{dsfont} 
\usepackage{indentfirst}
\usepackage{graphicx}
\usepackage[colorlinks]{hyperref}
\usepackage{cite}
\usepackage{colortbl}
\usepackage{physics}
\usepackage{microtype}
\usepackage[normalem]{ulem}
\usepackage{ulem}

\numberwithin{equation}{section}
\allowdisplaybreaks

\begin{document}

\baselineskip=18pt 
\baselineskip 0.6cm
\begin{titlepage}
\vskip 4cm

\begin{center}
\textbf{\LARGE Maxwell Chern-Simons gravity in 3D: Thermodynamics of cosmological solutions and black holes with torsion}
\par\end{center}{\LARGE \par}

\begin{center}
	\vspace{1cm}
    \textbf{Luis Avilés}$^{\ddag,\S}$,
    \textbf{Patrick Concha}$^{\dag, \star}$,
    \textbf{Javier Matulich}$^{\ast, \star}$,\\
	\textbf{Evelyn Rodríguez}$^{\dag, \star}$
    \textbf{David Tempo}$^{\ddag,\S}$,
	\small
	\\[5mm]
      $^{\ddag}$\textit{Instituto de Ciencias Exactas y Naturales (ICEN), Universidad Arturo Prat,}\\
     \textit{ Playa Brava 3256, 1111346 Iquique, Chile}
       \\[2mm]
        $^{\S}$\textit{Facultad de Ciencias, Universidad Arturo Prat,}\\
     \textit{Avenida Arturo Prat Chacón 2120, 1110939 Iquique, Chile}
       \\[2mm]
     $^{\dag}$\textit{Departamento de Matemática y Física Aplicadas, Universidad Católica de la Santísima Concepción, }\\
\textit{ Alonso de Ribera 2850, Concepción, Chile.}
  \\[2mm]
  $^{\star}$\textit{Grupo de Investigación en Física Teórica, GIFT, Universidad Católica de la Santísima Concepción, }\\
\textit{Alonso de Ribera 2850, Concepción, Chile.}
      \\[2mm]
	$^{\ast}$\textit{Instituto de Física Teórica UAM/CSIC,}\\
	\textit{C/ Nicolás Cabrera 13-15, Universidad Autónoma de Madrid, Cantoblanco, Madrid 28049,   Spain. }
 \\[5mm]
	\footnotesize
         \texttt{luaviles@unap.cl},
	\texttt{patrick.concha@ucsc.cl},
        \texttt{javier.matulich@csic.es},
	\texttt{erodriguez@ucsc.cl}
        \texttt{jtempol@unap.cl}
    
	\par\end{center}
\vskip 26pt
\centerline{{\bf Abstract}}
\medskip
\noindent  
 We construct generalized sets of asymptotic conditions for both three-dimensional Maxwell Chern–Simons gravity and a novel extension that incorporates torsion through a deformation of the Maxwell algebra. These boundary conditions include the most general temporal components of the gauge fields that consistently preserve the corresponding asymptotic Maxwell algebras with identical classical central charges, while allowing for the inclusion of chemical potentials conjugate to the conserved charges. We show that both sets of asymptotic configurations admit nontrivial solutions carrying not only mass and angular momentum but also an additional global spin-2 charge. In the torsionless case, the theory admits locally flat cosmological spacetimes, whereas in the presence of torsion, it generalizes to BTZ-like black hole geometries. For each case, the thermodynamic properties are consistently derived in terms of the gauge fields and the topology of the Euclidean manifold, shown to correspond to a solid torus. Furthermore, we obtain a general expression for the entropy, depending on both the horizon area and its spin-2 analogues, which can be written as a reparametrization-invariant integral of the induced spin-2 fields on the spacelike section of the horizon.

\end{titlepage}\newpage {\baselineskip=12pt \tableofcontents{}}

\section{Introduction}\label{sec1}

Three-dimensional gravity provides a fertile setting for exploring
conceptual and structural aspects of gravitational theories. Its topological
nature and the absence of local degrees of freedom simplify the dynamics,
allowing for the explicit construction of exact solutions and a detailed
analysis of their physical properties. Notably, when formulated as
a Chern-Simons (CS) gauge theory, gravity in three dimensions provides
a natural frame in which symmetries, topology, and geometry are intrinsically
linked \cite{Zanelli:2005sa}. The three-dimensional CS gravity framework
constitutes a powerful theoretical laboratory for exploring various
aspect of higher-dimensional gravity theories, including black hole
solutions and their associated thermodynamics properties \cite{Banados:1992wn,Banados:1992gq,Banados:1993ur,Crisostomo:2000bb}.

The inclusion of additional gauge symmetries has proven to be a fruitful
avenue for extending General Relativity (GR). In particular, the formulation
of CS (super)gravity theories through extensions, expansions, and
deformations of the Poincaré and AdS symmetries has led to significant
results \cite{Edelstein:2006se,Izaurieta:2009hz,Salgado:2014jka}.
Within this framework, the Maxwell algebra \cite{Bacry:1970ye,schrader1972maxwell,Gomis:2017cmt},
which corresponds to an extension and deformation of the Poincaré
algebra \cite{Safari:2020pje}, has attracted considerable attention
and has been studied in various contexts \cite{Duval:2008tr,deAzcarraga:2010sw,Durka:2011nf,deAzcarraga:2012qj,Concha:2013uhq,Concha:2014vka,Hoseinzadeh:2014bla,Concha:2014tca,Caroca:2017onr,Caroca:2017izc,Ravera:2018vra,Concha:2018zeb,Concha:2018jxx,Concha:2018ywv,Concha:2019eip,Salgado-Rebolledo:2019kft,Chernyavsky:2020fqs,Adami:2020xkm,Caroca:2021bjo,Concha:2024rac,Hofenstock:2025xoc}.
Its incorporation into the CS formulation of gravity give rise to
Maxwell CS gravity \cite{Salgado:2014jka}, which is characterized
by a new gauge field denoted as the gravitational Maxwell field\footnote{The CS theory based on Maxwell algebra in $2+1$ was initially considered in \cite{Cangemi:1992ri, Duval:2008tr} as a notable  model leading to the two-dimensional linear gravities referred in   \cite{Fukuyama:1985gg,Isler:1989hq,Chamseddine:1989yz,Cangemi:1992bj,Cangemi:1993sd} by a dimensional reduction.}. The
corresponding field equations derived from the CS action comprise
the Poincaré ones, describing asymptotically flat and torsionless
geometries, together with an additional equation involving the gravitational
Maxwell field.

The Maxwell CS gravity theory can alternatively be obtained as a vanishing
cosmological constant limit from two different gravity theories: the
so-called AdS-Lorentz gravity \cite{Soroka:2006aj,Salgado:2014qqa,Concha:2018jjj},
whose underlying symmetry corresponds to a semi-simple extension of
the Poincaré algebra, and the Maxwell with torsion \cite{Adami:2020xkm},
which is invariant under the deformed Maxwell algebra introduced
in \cite{Concha:2019eip}. Although both symmetries allow the inclusion
of a negative cosmological constant in the CS actions, the resulting
gravity theories are physically distinct. Specifically, Maxwell CS
with torsion can be regarded as a Maxwellian extension of Teleparallel
gravity \cite{Blagojevic:2003vn,Blagojevic:2003uc,Giacomini:2006dr,Caroca:2021njq}
when the deformation parameter is fixed to $\varepsilon=-2/\ell$,
with $\ell$ related to the cosmological constant by $\Lambda=-\frac{1}{\ell^{2}}$.
Notably, under this identification, the cosmological constant effectively
acts as a source of torsion.

In the study of asymptotic symmetries, Maxwell CS gravity has been
shown to possess a modified asymptotic structure, leading to an enlargement
of the $\mathfrak{bms}_{3}$ algebra \cite{Ashtekar:1996cd,Barnich:2006av,Barnich:2010eb},
commonly denoted as $\mathfrak{max}$-$\mathfrak{bms}_{3}$ \cite{Concha:2018zeb}.
This extended $\mathfrak{bms}_{3}$ algebra can also be recovered
as a flat limit of three copies of the Virasoro algebra \cite{Caroca:2017onr,Concha:2018jjj}\footnote{Also described as a semi-simple enlargement of the $\mathfrak{bms}_{3}$
algebra.}, which appears as the asymptotic symmetry algebra of the AdS-Lorentz
CS gravity \cite{Concha:2018jjj}. Moreover, when torsion is switched
on within the Maxwell CS gravity framework, the asymptotic symmetry
algebra undergoes an infinite-dimensional enhancement, taking the
form of $\mathfrak{bms}_{3}\oplus\mathfrak{vir}$, which can be interpreted
as an extended version of the deformed Maxwell algebra \cite{Concha:2019eip}.

Despite its algebraic richness and theoretical appeal, the thermodynamic
properties of Maxwell solutions have remained largely unexplored.
In this work, we address this gap by analyzing the entropy and thermodynamic
consistency of asymptotically flat cosmological solutions in three-dimensional
Maxwell CS theory. We show that the gravitational Maxwell field modifies the structure of canonical generators such that solutions are endowed not only with mass and angular momentum but also with an additional global spin-2 charge, which contributes to the first law of thermodynamics, thus providing new insights into the structure of the theory. In particular,
the entropy we obtain extends the standard Bekenstein--Hawking formula
of general relativity reported in \cite{Barnich:2012xq,Bagchi:2012xr}. We then
consider an extension of the theory by including non-vanishing torsion,
achieved via a deformation of the Maxwell algebra that leads to a
new CS theory of gravity with non-Riemannian geometry. As we have
said before, this torsional theory possesses a modified asymptotic
symmetry algebra and admits a novel black hole solution, which we
construct explicitly. We compute its conserved charges and entropy,
demonstrating that the solution is a natural generalization of the
BTZ-like black hole with torsion \cite{Garcia:2003nm,Blagojevic:2006jk,Blagojevic:2006nf,Blagojevic:2006hh,Blagojevic:2013aaa}.
Importantly, we show that the previously discussed cosmological solution
is recovered as the limit $\varepsilon\rightarrow0$ of this black
hole configuration. This limit can be interpreted as a vanishing cosmological
constant limit, i.e. $\ell\rightarrow\infty$, when we set $\varepsilon=-2/\ell$.     
To the best of our knowledge, this is the first explicit construction
of a black hole solution with torsion in Maxwell\footnote{A complete study of the stationary black hole solution, along with
its thermodynamic properties, of a CS action based on the semi-simple
extension of the Poincaré gauge group (or AdS-Lorentz) was considered
in \cite{Aviles:2023igk}.}, along with a complete thermodynamic analysis in the torsional and
torsionless sectors.

This paper is organized as follows. In Section 2, we review Maxwell
CS gravity and study the thermodynamics of its asymptotically flat
cosmological solutions. Section 3 introduces the torsional extension
and analyzes its asymptotic symmetries and black hole thermodynamics.
Section 4 concludes our work with a discussion of our results and
future directions.

\section{Three-dimensional Maxwell Chern-Simons gravity}

\label{sec2} In this section, we briefly review the so-called Maxwell
CS gravity theory formulated using the Chern-Simons formalism. We
review the asymptotic structure of the theory, whose algebra is given
by an extension and deformation of the $\mathfrak{bms}_{3}$ algebra
\cite{Concha:2018zeb}, and is denoted here by $\mathfrak{max}$-$\mathfrak{bms}_{3}$.
We also address the explicit computation of the entropy for an asymptotically
flat cosmological solution. The Maxwell algebra generated by $\{J_{a},P_{a},Z_{a}\}$
is given by the following non-vanishing commutators: 
\begin{eqnarray}
\left[J_{a},J_{b}\right]=\epsilon_{abc}J^{c}\,,\qquad\ \ \ \left[J_{a},P_{b}\right]=\epsilon_{abc}P^{c}\,,\nonumber \\
 \left[J_{a},Z_{b}\right]=\epsilon_{abc}Z^{c}\,,\qquad\ \ \ \left[P_{a},P_{b}\right]=\epsilon_{abc}Z^{c}\,.\label{Max-Alg}
\end{eqnarray}
where $a,b,\dots=0,1,2$ are Lorentz indices raised and lowered with
the Minkowski metric $\eta_{ab}$ and $\epsilon_{abc}$ is the Levi-Civita
tensor. In order to write down a CS action for this algebra, we define
the one-form gauge connection 
\begin{equation}
A=\omega^{a}J_{a}+e^{a}P_{a}+\sigma^{a}Z_{a}\,,\label{GaugeField-Max}
\end{equation}
where $e^{a},\omega^{a}$ and $\sigma^{b}$ correspond to the vielbein,
spin connection, and gravitational Maxwell field, respectively.

The non-vanishing components of the invariant tensor are given by:
\begin{align}
\left\langle J_{a}J_{b}\right\rangle  & =\alpha_{0}\eta_{ab}\,, & \left\langle P_{a}P_{b}\right\rangle  & =\alpha_{2}\eta_{ab}\,,\nonumber \\
\left\langle J_{a}P_{b}\right\rangle  & =\alpha_{1}\eta_{ab}\,, & \left\langle J_{a}Z_{b}\right\rangle  & =\alpha_{2}\eta_{ab}\,,\label{it1}
\end{align}
where $\alpha_{0}$, $\alpha_{1}$ and $\alpha_{2}$ are real dimensionless
constants. Then, considering the previous invariant tensor and the
one-form gauge connection (\ref{GaugeField-Max}) in the CS action
\begin{equation}
I[A]=\frac{k}{4\pi}\int_{\mathcal{M}}\left\langle AdA+\frac{2}{3}A^{3}\right\rangle \,,\label{CSaction}
\end{equation}
defined on a three-dimensional manifold $\mathcal{M}$, and where
$k=\frac{1}{4G}$ is the level of the theory related to the gravitational
constant $G$, we obtain 
\begin{eqnarray}
I_{\mathfrak{max}} & = & \frac{k}{4\pi}\int\alpha_{0}\left(\omega^{a}d\omega_{a}+\frac{1}{3}\epsilon_{abc}\omega^{a}\omega^{b}\omega^{c}\right)+2\alpha_{1}R_{a}e^{a}+\alpha_{2}\left(e^{a}T_{a}+2\sigma^{a}R_{a}\right)\,.\label{I-Max}
\end{eqnarray}
Here $R^{a}=d\omega^{a}+\frac{1}{2}\epsilon^{abc}\omega_{b}\omega_{c}$
and $T^{a}=de^{a}+\epsilon^{abc}\omega_{b}e_{c}$, are the usual Lorentz
curvature and torsion two-forms. The first term along $\alpha_{0}$
in the action \eqref{I-Max} is the gravitational CS term. Next term
with coupling constant $\alpha_{1}$ is the Einstein-Hilbert term.
The last term, with the strength of interaction $\alpha_{2}$, gives
the dynamics to the gravitational Maxwell field and also contributes
to the dynamics of other fields. Indeed, when $\alpha_{2}\neq0$,
the field equations are given by 
\begin{equation}
\begin{split}R^{a} & =0\,,\\
T^{a} & =0\,,\\
D\sigma^{a}+\frac{1}{2}\epsilon^{abc}e_{b}e_{c} & =0\,,
\end{split}
\label{eq:Eqs.M-Max}
\end{equation}
where $D\Phi^{a}=d\Phi^{a}+\epsilon_{\,\,\,bc}^{a}\omega^{b}\Phi^{c}$
denotes the Lorentz covariant derivative. These field equations describe
geometries which are Riemannian and locally flat, as it can be seen
from the above first two expressions. The third equation shows the
dynamics of the gravitational Maxwell field $\sigma^{a}$, referred
here as a ``spin-2'' field \cite{Chernyavsky:2020fqs}\footnote{Here, the statement that the gravitational Maxwell field  $\sigma^{a}$ describes a “spin-2” field is based on the fact that in the metric formulation only the symmetric part remains so that its field equation reduces to the massless Fierz-Pauli equation.}.
As it was shown in \cite{Concha:2018zeb} the coupling of this one-form
field to the geometry leads to nontrivial effects compared to GR that
can be regarded as a deformation \cite{Chernyavsky:2020fqs} of the
“exotic” Einstein gravity \cite{Witten:1988hc}. In what follows the
asymptotic structure of the three-dimensional Maxwell CS gravity is
explored including the thermodynamics features of locally flat cosmological
spacetimes.

\subsection{Asymptotic conditions with chemical potentials}

\label{subsec:AsympCond-Max}

Here we review and generalize the set of boundary conditions for Maxwell
CS gravity in \cite{Concha:2018zeb} to incorporate chemical potentials
conjugated to the conserved charges. The study of the behavior of
dynamic fields in the asymptotic region become crucial for computing
the thermodynamic properties configurations with a sensible thermodynamics.

As explained in \cite{Coussaert:1995zp} the radial dependence of
the gauge field can be entirely gauged away asymptotically by virtue
of a suitable choice of a gauge group element $g=g\left(r\right)$,
so that $A=g^{-1}ag+g^{-1}dg$, where $a$ is an auxiliary connection
given by 
\begin{equation}
a=a_{\phi}\left(t,\phi\right)d\phi+a_{t}\left(t,\phi\right)dt\;,\label{eq:a}
\end{equation}
The asymptotic behavior for $a_{\phi}$ can be written as follows
\cite{Concha:2018zeb,Matulich:2023xpw}\footnote{Hereafter, our conventions are such that the nonvanishing components
of the Minkowski metric in tangent space $\eta_{ab}$ read $\eta_{01}=\eta_{10}=\eta_{22}=1$,
and the Levi-Civita symbol fulfills $\epsilon_{012}=1$.} 
\begin{equation}
a_{\phi}=J_{1}+\frac{1}{2}{\cal \,M}\,J_{0}+\frac{1}{2}{\cal N}\,P_{0}+\frac{1}{2}\mathcal{F}\,Z_{0}\,,\label{a_phi-Max}
\end{equation}
where $\mathcal{N},\mathcal{M}$, and $\mathcal{F}$ stand for arbitrary
functions of the boundary coordinates $(t,\phi)$. The asymptotic
form of $a_{\phi}$ is preserved under a restricted set of gauge transformations,
$\delta a=d\lambda+[a,\lambda]$, with the Lie-algebra-valued parameter
$\lambda_{(0)}=\lambda_{(0)}[y,f,h]$ generated by 
\begin{align}
\lambda_{(0)}[y,f,h] & =yJ_{1}-y^{\prime}J_{2}+fP_{1}-f^{\prime}P_{2}+hZ_{1}-h^{\prime}Z_{2}+\left(\frac{\mathcal{M}}{2}f+\frac{\mathcal{N}}{2}y-f^{\prime\prime}\right)P_{0}\nonumber \\
 & +\left(\frac{\mathcal{M}}{2}y-y^{\prime\prime}\right)J_{0}+\left(\frac{1}{2}\mathcal{M}h+\frac{1}{2}\mathcal{F}y+\frac{1}{2}\mathcal{N}f-h^{\prime\prime}\right)Z_{0}\,,\label{eq:lambda-Max}
\end{align}
provided that the functions $\mathcal{M},\mathcal{N}$ and $\mathcal{F}$
transform according to: 
\begin{align}
\delta_{(0)}{\cal M} & ={\cal M}^{\prime}y+2{\cal M}y^{\prime}-2y{}^{\prime\prime\prime}\,,\nonumber \\
\delta_{(0)}{\cal N} & ={\cal M}^{\prime}f+2{\cal M}f^{\prime}-2f{}^{\prime\prime\prime}+{\cal N}^{\prime}y+2{\cal N}y^{\prime}\,,\label{TransfLaw-MNF-Max}\\
\delta_{(0)}{\cal F} & ={\cal M}^{\prime}h+2{\cal M}h^{\prime}-2h{}^{\prime\prime\prime}+{\cal N}^{\prime}f+2{\cal N}f^{\prime}+{\cal F}^{\prime}y+2{\cal F}y^{\prime}\,,\nonumber 
\end{align}
where $y=y(\phi,t)$, $f=f(\phi,t)$, and $h=h(\phi,t)$ are arbitrary
functions and prime denotes derivative with respect to $\phi$. Following
Refs. \cite{Henneaux:2013dra,Bunster:2014mua}, the asymptotic symmetries
are preserved under time evolution by choosing the asymptotic form
of the Lagrange multiplier $a_{t}$ according to 
\begin{equation}
a_{t}=\lambda_{(0)}[\mu,\xi,\nu]\,,\label{eq:at-Max}
\end{equation}
with $\lambda$ defined through \eqref{eq:lambda-Max} and where $\mu,\xi$,
and $\nu$ stand for arbitrary functions of the boundary coordinates
$(t,\phi)$ which are assumed to be fixed at the boundary. In the
thermodynamic analysis these functions are identified with the ``chemical
potentials'' conjugated to the charges \cite{Henneaux:2013dra}.
The asymptotic form of the temporal components of the connection in
\eqref{eq:at-Max} is maintained by the asymptotic symmetries provided
both that the field equations are fulfilled in the asymptotic region,
i.e., 
\begin{align}
\dot{\mathcal{M}} & =2\mathcal{M}\mu^{\prime}+\mathcal{M}^{\prime}\mu-2\mu^{\prime\prime\prime}\,,\nonumber \\
\dot{\mathcal{N}} & =2\mathcal{M}\xi^{\prime}+\mathcal{M}^{\prime}\xi+2\mathcal{N}\mu^{\prime}+\mathcal{N}^{\prime}\mu-2\xi^{\prime\prime\prime}\,,\label{eq:EqsMotion-MNF-Max}\\
\dot{\mathcal{F}} & =2\mathcal{M}\nu^{\prime}+\mathcal{M}^{\prime}\nu+2\mathcal{N}\xi^{\prime}+\mathcal{N}^{\prime}\xi+2\mathcal{F}\mu^{\prime}+\mathcal{F}^{\prime}\mu-2\nu^{\prime\prime\prime}\,,\nonumber 
\end{align}
and that parameters $\mu,\xi$, and $\nu$, which describe the asymptotic
symmetries, satisfy suitable differential equations of first order
in time given by 
\begin{align}
\dot{y} & =y\mu^{\prime}-\mu y^{\prime}\quad;\quad\dot{\xi}=f\mu^{\prime}-\mu f^{\prime}+y\xi^{\prime}-\xi y^{\prime}\;,\nonumber \\
\dot{\nu} & =h\mu^{\prime}-h^{\prime}\mu+f\xi^{\prime}-f^{\prime}\xi+y\nu^{\prime}-y^{\prime}\nu\;.
\end{align}
The asymptotic symmetry generators can be computed in the Regge-Teitelboim
approach \cite{REGGE1974286,Banados:1994tn} such that their variations
reduce to 
\begin{equation}
\delta Q[\lambda]=-\frac{k}{2\pi}\int\limits_{\partial\Sigma}\left\langle \lambda\delta a\right\rangle d\phi\,,\label{eq:deltaQ}
\end{equation}
where $\partial\Sigma$ stands for the boundary of the spacelike section
$\Sigma$. The explicit form of the generators associated to the asymptotic
symmetries can be integrated and are found to be given by 
\begin{equation}
Q[y,f,h]=-\int d\phi\left(y\mathcal{J}+f\mathcal{P}+h{\cal Z}\right)\,,\label{Q-Max}
\end{equation}
where the dynamical fields $\mathcal{J}$, $\mathcal{P}$ and $\mathcal{Z}$
are determined in terms of the functions $\mathcal{N},\mathcal{M}$,
and $\mathcal{F}$, according to the relations: 
\begin{equation}
\mathcal{J}=\frac{k}{4\pi}\,\left(\alpha_{2}{\cal F}+\alpha_{1}{\cal N}+\alpha_{0}{\cal M}\right)\quad;\quad\mathcal{P}=\frac{k}{4\pi}\,\left(\alpha_{2}{\cal N}+\alpha_{1}{\cal M}\right)\quad;\quad\mathcal{Z}=\frac{k}{4\pi}\,\alpha_{2}{\cal M}\,.\label{eq:AuxVar-Max}
\end{equation}
The transformations law of the dynamical variables $\{\mathcal{J},\mathcal{P},\mathcal{Z}\}$
are found using \eqref{eq:AuxVar-Max} and \eqref{TransfLaw-MNF-Max}
and they explicitly read 
\begin{align}
\delta\mathcal{J} & =2{\cal J}y^{\prime}+{\cal J}^{\prime}y+2{\cal P}f^{\prime}+{\cal P}^{\prime}f+2{\cal Z}h^{\prime}+{\cal Z}^{\prime}h-\frac{\alpha_{0}k}{2\pi}y{}^{\prime\prime\prime}-\frac{\alpha_{1}k}{2\pi}f{}^{\prime\prime\prime}-\frac{\alpha_{2}k}{2\pi}h{}^{\prime\prime\prime}\,,\nonumber \\
\delta\mathcal{P} & =2{\cal P}y^{\prime}+{\cal P}^{\prime}y+2{\cal Z}f^{\prime}+{\cal Z}^{\prime}f-\frac{\alpha_{1}k}{2\pi}y{}^{\prime\prime\prime}-\frac{\alpha_{2}k}{2\pi}f{}^{\prime\prime\prime}\,,\label{eq:TransfLaw-JPZ-Max}\\
\delta\mathcal{Z} & =2{\cal Z}y^{\prime}+{\cal Z}^{\prime}y-\frac{\alpha_{2}k}{2\pi}y{}^{\prime\prime\prime}\,.\nonumber 
\end{align}
The algebra of the conserved charges \eqref{Q-Max} can be readily
obtained from the transformation law of the dynamical fields in \eqref{eq:TransfLaw-JPZ-Max}
by virtue of $\delta_{\lambda_{2}}Q[\lambda_{1}]=\left\{ Q[\lambda_{1}],Q[\lambda_{2}]\right\} $.
Thus, expanding in Fourier modes according to $X=\frac{1}{2\pi}\sum X_{m}e^{im\phi}$,
and providing that the zero modes $X_{0}$ being shifted as $X_{0}\rightarrow X_{0}+\frac{c_{i}}{24}$,
the nontrivial Poisson brackets are given by\cite{Concha:2018zeb,Matulich:2023xpw}
\begin{align}
i\left\{ \mathcal{J}_{m},\mathcal{J}_{n}\right\}  & =\left(m-n\right)\mathcal{J}_{m+n}+c_{{\cal J}}\,m(m^{2}-1)\delta_{m+n,0}\,,\nonumber \\
\,i\left\{ \mathcal{J}_{m},\mathcal{P}_{n}\right\}  & =\left(m-n\right)\mathcal{P}_{m+n}+c_{{\cal P}}\,m(m^{2}-1)\delta_{m+n,0}\,,\nonumber \\
i\left\{ \mathcal{J}_{m},\mathcal{Z}_{n}\right\}  & =\left(m-n\right)\mathcal{Z}_{m+n}+c_{{\cal Z}}\,m(m^{2}-1)\delta_{m+n,0}\,,\label{eq:MaxAlgModes}\\
i\left\{ \mathcal{P}_{m},\mathcal{P}_{n}\right\}  & =\left(m-n\right)\mathcal{Z}_{m+n}+c_{{\cal Z}}\,m(m^{2}-1)\delta_{m+n,0}\,,\nonumber 
\end{align}
where the central extensions $c_{{\cal J}}$, $c_{{\cal P}}$ and
$c_{{\cal Z}}$ are fully determined in terms of the constants of
the action \eqref{I-Max} according to 
\begin{equation}
c_{{\cal J}}=k\alpha_{0}\,\qquad c_{{\cal P}}=k\alpha_{1}\,,\qquad c_{{\cal Z}}=k\alpha_{2}\,.\label{centralc}
\end{equation}
It is worth noting that this asymptotic symmetry algebra corresponds
to an extension and deformation of $\mathfrak{bms}_{3}$ algebra \cite{Ashtekar:1996cd,Barnich:2006av,Barnich:2010eb}
at which both the abelian generators $\mathcal{Z}_{n}$ and the generators
$\mathcal{P}_{n}$ have conformal weight $2$. This deformation introduces
an additional central extension $c_{{\cal Z}}$ which is proportional
to the coupling constant of the Maxwell gravitational action. It also
is reassuring that the asymptotic symmetry algebra \eqref{eq:MaxAlgModes}
contains the wedge algebra in \eqref{Max-Alg} for generators $X_{m}$,
with $m=-1,0,1$, providing $i\{,\}\rightarrow[,]$.

It should be noted that gauge field configurations with $\mathcal{M}$,
$\mathcal{N}$, $\mathcal{F}$, as well as the chemical potentials
$\xi$, $\mu$, $\nu$ fixed to constants, solve the field equations
in \eqref{eq:EqsMotion-MNF-Max}. Indeed, this class of configurations
explicitly reads 
\begin{align}
a & =\left(J_{1}+\frac{1}{2}{\cal \,M}\,J_{0}+\frac{1}{2}{\cal N}\,P_{0}+\frac{1}{2}\mathcal{F}\,Z_{0}\right)d\phi+\left[\mu J_{1}+\frac{\mu}{2}\mathcal{M}J_{0}\right.\nonumber \\
 & \left.+\frac{1}{2}\left(\xi\mathcal{M}+\mu\mathcal{N}\right)P_{0}+\xi P_{1}+\frac{1}{2}\left(\nu\mathcal{M}+\mu\mathcal{F}+\xi\mathcal{N}\right)Z_{0}+\nu Z_{1}\right]dt\,,\label{eq:a-Cosmo-Mx}
\end{align}
In appendix \ref{sec:MetricF-Max} by restoring the radial dependence
of the gauge fields, it shown that this stationary solution describes
a locally flat cosmological spacetime endowed with a spin-2 gauge
field, named here as the Maxwell CS field. It also follows from \eqref{Q-Max}
that this solution not only possesses mass and angular momentum, being
parametrized by the spin-2 charges ${\cal P}$, ${\cal J}$, respectively,
but also has an extra global charge of spin 2, determined by ${\cal Z}$.
In the next section thermodynamic properties of this solution is analyzed.

\medskip{}

\subsection{Thermodynamics of asymptotically flat cosmological solutions}\label{sec:Max-Termo}

This section is devoted to the study of the thermodynamics of cosmological
solutions in the Maxwell CS gravity theory. In this context, following
the canonical approach the entropy can be suitably computed along
the lines of \cite{Matulich:2014hea,Bunster:2014mua}, by the general
formula 
\begin{align}
S & =\frac{k}{2\pi}\left[\int_{r_{+}}d\tau d\phi\langle A_{\tau}A_{\phi}\rangle\right]_{\text{on-shell}}\;\nonumber \\
 & =k\left[\langle a_{\tau}a_{\phi}\rangle\right]_{\text{on-shell}}\;,\label{Entropy-CS}
\end{align}
where $A_{\tau}$ stands for the Euclidean continuation of $A_{t}$
by the replacement $\tau=it$ and where its final form is obtained
in terms of the on-shell holonomies which is then continued to Lorentzian.

For the field configuration in \eqref{eq:a-Cosmo-Mx} the entropy
reads

\begin{equation}
S=k\left[\alpha_{0}\mu\mathcal{M}+\alpha_{1}\left(\mu\mathcal{N}+\xi\mathcal{M}\right)+\alpha_{2}\left(\nu\mathcal{M}+\xi\mathcal{N}+\mu\mathcal{F}\right)\right]\,,\label{eq:Entropy-OfSh-Max}
\end{equation}
where it is assumed that the chemical potentials $(\xi,\mu,\nu)$
are constrained to fulfill regularity conditions. Since all the chemical
potentials are explicitly incorporated along the temporal components
of the gauge fields, the analysis can be carried out for a fixed range
of the angular coordinates of the torus, i.e., we assume that $0<\tau\leq1$,
and $0<\phi\leq2\pi$. Thus, these regularities conditions, can be
computed following the prescription in \cite{Matulich:2014hea} which,
in our case consists in finding a permissible gauge transformation
that allows to gauge away the temporal components of the dreibein
and the gravitational Maxwell field, i.e, $e_{t}=\sigma_{t}=0$. The
latter implies that $a_{t}$ reduces such that it only takes values
in the Lorentz subalgebra of the gauge group spanned by the Maxwell
algebra \eqref{Max-Alg}, so that the remaining regularity condition
results in requiring that the holonomy along the thermal circle to
be trivial.

In the present case, the required group element reads 
\begin{equation}
g=e^{\lambda P_{2}+\rho Z_{2}}\;,
\end{equation}
so that the angular and temporal components of the gauge field now
read 
\begin{align}
a_{\phi} & =J_{1}+\frac{1}{2}\mathcal{M}J_{0}+\lambda P_{1}+\left(\frac{\lambda^{2}}{2}+\rho\right)Z_{1}+\frac{1}{2}\left(\mathcal{N}-\lambda\mathcal{M}\right)P_{0}+\frac{1}{2}\left[\mathcal{F}+\left(\frac{\lambda^{2}}{2}-\rho\right)\mathcal{M}-\lambda\mathcal{N}\right]Z_{0}\,,\nonumber \\
a_{t} & =\mu\left(J_{1}+\frac{1}{2}\mathcal{M}J_{0}\right)+(\xi+\mu\lambda)P_{1}+\left[\nu+\xi\lambda+\frac{1}{2}\mu(\lambda^{2}+4\rho)\right]Z_{1}+\frac{1}{2}\left(\xi\mathcal{M}+\mu\mathcal{N}-\mu\lambda\mathcal{M}\right)P_{0}\nonumber \\
 & +\frac{1}{4}\left[2\mu\mathcal{F}+\left(2\nu-\xi\lambda+\mu\lambda^{2}-4\mu\rho\right)\mathcal{M}+2\left(\xi-\mu\lambda\right)\mathcal{N}\right]Z_{0}\,,\label{eq:at-g}
\end{align}
from which it follows that the contributions along the generators
$P_{1}$ and $Z_{1}$ in \eqref{eq:at-g} can be removed provided
that $\lambda$ and $\rho$ are chosen as follow 
\begin{equation}
\lambda=-\frac{\xi}{\mu}\,,\quad\rho=\frac{1}{2}\frac{\xi^{2}}{\mu^{2}}-\frac{\nu}{\mu}\,.
\end{equation}
This subsequently allows for fixing the Lagrange multiplier parameters
$\mu$ and $\nu$ (chemical potentials) according to, 
\begin{equation}
\mu=-2\frac{\mathcal{M}}{\mathcal{N}}\xi\,,\qquad\nu=\left(\frac{\mathcal{F}}{\mathcal{N}}-\frac{3}{4}\frac{\mathcal{N}}{\mathcal{M}}\right)\xi\,,
\end{equation}
the terms along $P_{0}$ and $Z_{0}$ in the temporal component of
the gauge field in \eqref{eq:at-g} vanish so that 
\begin{align}
a_{t} & =-2\xi\frac{\mathcal{M}}{\mathcal{N}}\left(J_{1}+\frac{1}{2}\mathcal{M}J_{0}\right)\,.
\end{align}
Hence, imposing that the holonomy of the spin connection around the
thermal circle is trivial, 
\begin{equation}
H=e^{a_{\tau}}\Big|_{\text{on-shell}}=(-1)^{n_{\mathfrak{m}}}\mathds{1}_{2\times2}\;,
\end{equation}
in the fundamental representation of $sl(2,\mathbb{R})$, this requirement
leads to 
\begin{equation}
\xi=-\frac{n_{\mathfrak{m}}\pi\mathcal{N}}{\mathcal{M}^{3/2}}\,,
\end{equation}
where $n_{\mathfrak{m}}$ is an integer. It is important to note that,
since we are dealing with a cosmological horizon, the orientation
of the solid torus is reversed relative to that of the black hole.
As a result, the chemical potential $\xi$ is such that $\xi=-1/T_{C}$.
This choice is consistent with the requirement that the Hawking temperature
be positive, fixing $n_{\mathfrak{m}}$ to be positive or negative,
depending on the sign of the function $\mathcal{N}$.

In summary, the regularity conditions of the Euclidean gauge fields
are successfully implemented, leading to the chemical potentials being
fixed as specific functions of the global charges, according to \eqref{eq:AuxVar-Max},
as follows 
\begin{align}
\xi & =-\frac{n_{\mathfrak{m}}\pi\mathcal{N}}{\mathcal{M}^{3/2}}\,, & \mu & =\frac{2\pi n_{\mathfrak{m}}}{\mathcal{M}^{1/2}}\,, & \nu & =-n_{\mathfrak{m}}\pi\left[\frac{\mathcal{F}}{\mathcal{M}^{3/2}}-\frac{3\mathcal{N}^{2}}{4\mathcal{M}^{5/2}}\right]\,.\label{eq:RegCond-Hol-Max}
\end{align}
It is reassuring that for $n_{\mathfrak{m}}=\text{sgn}({\cal N})$
the above result for the regularity conditions is in complete agreement
with those found in the metric formalism (see appendix
\ref{sec:MetricF-Max}), which stem from requiring
the absence of conical singularities of the Euclidean fields at the
cosmological horizon, as it is shown in \eqref{eq:RegCond-Max-MetricF}. In fact, it is found that the existence of the cosmological horizon implies that $\mathcal{M}>0$, and that the Euclidean fields possess around the horizon the  topology  of a solid torus leads to the condition:
\begin{equation}
\frac{\mathcal{N}}{\mathcal{M}}\left(\mathcal{F}-\frac{\mathcal{N}^{2}}{4\mathcal{M}}\right)>0\;.\label{Max-SameTopo}
\end{equation}
The entropy can then be readily obtained by plugging the chemical
potentials \eqref{eq:RegCond-Hol-Max} in eq.\eqref{eq:Entropy-OfSh-Max},
so in terms of the variables $\mathcal{M}$, $\mathcal{N}$ and $\mathcal{F}$,
it reads 
\begin{equation}
S=2\pi n_{\mathfrak{m}}k\left[\alpha_{0}\sqrt{\mathcal{M}}+\alpha_{1}\frac{\mathcal{N}}{2\sqrt{\mathcal{M}}}+\frac{\alpha_{2}}{2}\left(\frac{\mathcal{F}}{\sqrt{\mathcal{M}}}-\frac{\mathcal{N}^{2}}{4\mathcal{M}^{3/2}}\right)\right]\,.\label{Entropy-Max}
\end{equation}
This result for the entropy not only extends the one for General Relativity
(along $\alpha_{1}$) \cite{Barnich:2012xq,Bagchi:2012xr,Matulich:2014hea}
in the presence of the CS term ($\alpha_{0}\neq0$) \cite{Bagchi:2013qva,Riegler:2014bia},
but also points out an explicit contribution (along $\alpha_{2}$)
of the gravitational Maxwell field to the entropy of the cosmological
spacetimes in this gravity theory.

In terms of the (extensive) global charges $\mathcal{J},\mathcal{P}$
and $\mathcal{Z}$ defined through eqs. \eqref{eq:AuxVar-Max}, the
entropy in \eqref{Entropy-Max} can be rewritten as follows 
\begin{equation}
S_{C}=2\pi n_{\mathfrak{m}}\sqrt{\pi k}\sqrt{\alpha_{2}}\left[\frac{\mathcal{J}}{\sqrt{\mathcal{Z}}}+\frac{\alpha_{0}}{\alpha_{2}}\sqrt{\mathcal{Z}}-\frac{1}{4}\left(\frac{\mathcal{P}}{\mathcal{Z}}-\frac{\alpha_{1}}{\alpha_{2}}\right)^{2}\sqrt{\mathcal{Z}}\right]\,.\label{S-MaxJPZ2}
\end{equation}
In order to have a sensible thermodynamics, the entropy must be a
real function, therefore ${\cal Z}>0$ as well as the coupling constants
$\alpha_{i}\geq0$ with $i=0,1$, $2$. In addition, positivity of
the entropy implies that the angular momentum has to be such that
for $n_{\mathfrak{m}}>0$ then $\mathcal{J}>\mathcal{J}_{c}$, and
for $n_{\mathfrak{m}}<0$ then $\mathcal{J}<\mathcal{J}_{c}$, where\footnote{In the metric formulation $n_{\mathfrak{m}}$ appears determined in
terms the global charges according to $n_{\mathfrak{m}}=\text{sgn}(\alpha_{2}\mathcal{P}-\alpha_{1}\mathcal{Z)}$.} 
\begin{equation}
\mathcal{J}_{c}=\frac{(\alpha_{2}\mathcal{P}-\alpha_{1}\mathcal{Z})^{2}-4\alpha_{0}\alpha_{2}\mathcal{Z}^{2}}{4\alpha_{2}^{2}\mathcal{Z}}\,.
\end{equation}

Thus, in order to determine the temperature and the chemical potentials
in the microcanonical ensemble, we use the thermodynamic relations
\begin{align}
\beta_{C}= & \left.\left(\frac{\partial S_{C}}{\partial\mathbb{M}}\right)\right|_{\mathbb{J},\mathbb{W}}=-T_{C}^{-1}=-n_{\mathfrak{m}}\frac{1}{2}\sqrt{\frac{\pi\alpha_{2}k}{\mathcal{Z}}}\left(\frac{\mathcal{P}}{\mathcal{Z}}-\frac{\alpha_{1}}{\alpha_{2}}\right)\,,\\
\Omega_{C}=- & \beta_{C}^{-1}\left.\left(\frac{\partial S_{C}}{\partial\mathbb{J}}\right)\right|_{\mathbb{M},\mathbb{W}}=-2\left(\frac{\mathcal{P}}{\mathcal{Z}}-\frac{\alpha_{1}}{\alpha_{2}}\right)^{-1}\,,\\
\Phi_{C}=- & \beta_{C}^{-1}\left.\left(\frac{\partial S_{C}}{\partial\mathbb{Z}}\right)\right|_{\mathbb{M},\mathbb{J}}=\frac{1}{4}\left(3\frac{\mathcal{P}}{\mathcal{Z}}+\frac{\alpha_{1}}{\alpha_{2}}\right)-\frac{\alpha_{2}\mathcal{J}-\alpha_{0}\mathcal{Z}}{\alpha_{2}\mathcal{P}-\alpha_{1}\mathcal{Z}}\,,
\end{align}
where $\mathbb{M}=2\pi\mathcal{P}$, $\mathbb{J}=-2\pi\mathcal{J}$,
and $\mathbb{W}=2\pi\mathcal{Z}$, such that the first law of thermodynamics
is found to fulfilled according to 
\begin{equation}
\delta S_{C}=\beta_{C}(\delta\mathbb{M}-\Omega_{C}\delta\mathbb{J}-\Phi_{C}\delta\mathbb{W})\,,
\end{equation}
such that the chemical potentials $\beta_{C}=\xi$, $\Omega_{C}=\mu/\xi$,
and $\Phi_{C}=-\nu/\xi$ correspond to the conjugated to the mass ($\mathbb{M}$),
angular momentum  ($\mathbb{J}$) and the additional spin-2 charge  ($\mathbb{W}$), respectively. It is important to emphasize that the mass of the cosmological solution is found to be bounded from below for ${\cal N}>0$, which turns out to be  compatible with the condition in \eqref{Max-SameTopo}.

As a final remark of this section, note that the entropy depends not
only on the horizon area but also on the values of the spin connection
and the gravitational Maxwell fields at the horizon (see appendix
\ref{sec:MetricF-Max}). Actually, one verifies that the values of
the purely angular components of the spin-2 fields at the cosmological
horizon are given by 
\begin{align}
\left(\frac{{\cal A}_{\text{metric}}}{2\pi}\right)^{2} & =e_{a\phi}e_{\;\phi}^{a}|_{r_{c}}=\frac{\mathcal{N}^{2}}{4\mathcal{M}}\;,\\
\left(\frac{{\cal A}_{\text{Max}}}{2\pi}\right)^{2} & =\sigma_{a\phi}\sigma_{\;\phi}^{a}|_{r_{c}}=\left(\frac{\mathcal{F}}{\sqrt{\mathcal{M}}}-\frac{\mathcal{N}^{2}}{4\mathcal{M}^{3/2}}\right)^{2}\;,\label{eq:Area-Analog-Max}\\
\left(\frac{{\cal A}_{\text{CS}}}{2\pi}\right)^{2} & =\omega_{a\phi}\omega_{\;\phi}^{a}|_{r_{c}}={\cal M}\;.
\end{align}
From this one concludes that the entropy \eqref{Entropy-Max} can
be fully expressed as the sum of the horizon area element and its
spin-2 analogues, thereby extending the standard Bekenstein-Hawking
formula for the entropy in our case to the form 
\begin{equation}
S=k\left(\alpha_{0}{\cal A}_{\text{CS}}+\alpha_{1}{\cal A}_{\text{metric}}+\alpha_{2}{\cal A}_{\text{Max}}\right),
\end{equation}
As depicted in appendix \ref{sec:MetricF-Max}, these particular contributions
in \eqref{eq:Area-Analog-Max} appear naturally defined in terms of
the pullback of the metric and the spin-2 fields at the spacelike
section of the horizon \cite{Perez:2013xi,Grumiller:2016kcp}.

In the following section, we will deal with the Maxwell generalization
of three-dimensional gravity with torsion. We first present the CS
gravity theory invariant under a deformed Maxwell algebra. Subsequently,
we will compute the asymptotic symmetry algebra considering consistent
boundary conditions that incorporate chemical potentials, and finally
we analyze the black hole solution of the theory including its corresponding
thermodynamics features. We will show that the results obtained in
this section can be recovered in the torsionless limit in a particular
case.

\section{Three-dimensional Maxwell Chern-Simons gravity with torsion}

\label{sec3}

In this section, we present a generalization of Maxwell CS gravity
by including torsion. For this purpose, we will consider a particular
deformation of the Maxwell algebra, which was first presented in \cite{Concha:2019eip},
and subsequently approached in \cite{Adami:2020xkm}. Interestingly,
despite the deformed Maxwell symmetry is isomorphic to the direct
sum of $\mathfrak{iso}(2,1)$ with $\mathfrak{so}(2,1)$, the resulting
CS action leads to a non-vanishing torsion as equation of motion \cite{Adami:2020xkm}.

\subsection{Maxwell Chern-Simons gravity with torsion}

The deformed Maxwell algebra is spanned by the set of generators $\{J_{a},P_{a},Z_{a}\}$,
which satisfy the following non-vanishing commutation relations: 
\begin{eqnarray}
 &  & \left[J_{a},J_{b}\right]=\epsilon_{abc}J^{c}\,,\qquad\ \ \ \left[J_{a},P_{b}\right]=\epsilon_{abc}P^{c}\,,\nonumber \\
 &  & \left[J_{a},Z_{b}\right]=\epsilon_{abc}Z^{c}\,,\qquad\ \ \left[P_{a},P_{b}\right]=\epsilon_{abc}\left(Z^{c}+\varepsilon P^{c}\right)\,.\label{MaxTor-Alg}
\end{eqnarray}
Note that when the deformation parameter $\varepsilon$ is fixed to
zero, the Maxwell algebra in \eqref{Max-Alg} is recovered. In fact,
the algebra \eqref{MaxTor-Alg} can be seen as the Maxwell extension
of the teleparallel algebra \cite{Caroca:2021njq}. As was pointed
out in \cite{Concha:2019eip}, the previous algebra can be written
as the direct sum of the $\mathfrak{iso}(2,1)$ and $\mathfrak{so}(2,1)$
algebras. Indeed, considering the following redefinition of the generators,
\begin{equation}
L_{a}\equiv J_{a}-\varepsilon^{-1}P_{a}-\varepsilon^{-2}Z_{a}\quad;\quad T_{a}\equiv\varepsilon^{-1}Z_{a}\quad;\quad S_{a}\equiv\varepsilon^{-1}P_{a}+\varepsilon^{-2}Z_{a}\,,\label{MaxTor-NewB-LTS}
\end{equation}
the $\mathfrak{iso}(2,1)\otimes\mathfrak{so}(2,1)$ algebra is revealed,
\begin{equation}
\begin{split} & [L_{a},L_{b}]=\epsilon_{abc}L^{c}\,,\\
 & [L_{a},T_{b}]=\epsilon_{abc}T^{c}\,,\\
 & [S_{a},S_{b}]=\epsilon_{abc}S^{c}\,.
\end{split}
\label{algebradirect}
\end{equation}
Three-dimensional gravity action based on the deformed Maxwell algebra
\eqref{MaxTor-Alg} can be formulated as a CS form by considering
\eqref{CSaction}, and where the gauge connection one-form $A$ reads
\begin{equation}
A=e^{a}P_{a}+\omega^{a}J_{a}+\sigma^{a}Z_{a}\,,\label{one-form2}
\end{equation}
where $e^{a}$, $\omega^{a}$ and $\sigma^{a}$ are the dreibein,
the spin connection, and the gravitational Maxwell field, respectively.
The corresponding curvature two-form $F=dA+\frac{1}{2}[A,A]$ is given
by 
\begin{equation}
F=\hat{T}^{a}P_{a}+R^{a}J_{a}+F^{a}Z_{a},\label{two-form2}
\end{equation}
with
\begin{align}
R^{a} & =d\omega^{a}+\frac{1}{2}\epsilon^{abc}\omega_{b}\omega_{c}\,,\nonumber \\
\hat{T}^{a} & =T^{a}+\frac{\varepsilon}{2}\epsilon^{abc}e_{b}e_{c}\,,\label{eq:Curvatures-MaxTor}\\
F^{a} & =D\sigma^{a}+\frac{1}{2}\epsilon^{abc}e_{b}e_{c}\,.\nonumber 
\end{align}
Here, $T^{a}=De^{a}$ is the torsion two-form, $R^{a}$ is the curvature
two-form, and $D$ is the Lorentz covariant derivative defined above.
Note that the limit $\varepsilon\rightarrow0$ reproduces, as expected,
the Maxwell curvature two-form in \eqref{eq:Eqs.M-Max}.

On the other hand, the non-vanishing components of an invariant bilinear
form of the deformed Maxwell algebra read 
\begin{equation}
\begin{split} & \langle J_{a}J_{b}\rangle=\alpha_{0}\eta_{ab}\,,\hspace{0.7cm}\langle P_{a}P_{b}\rangle=(\varepsilon\alpha_{1}+\alpha_{2})\eta_{ab}\,,\\
 & \langle J_{a}P_{b}\rangle=\alpha_{1}\eta_{ab}\,,\hspace{0.7cm}\langle J_{a}Z_{b}\rangle=\alpha_{2}\eta_{ab}\,,
\end{split}
\label{eq:invtensor2}
\end{equation}
where $\alpha_{0},\alpha_{1}$ and $\alpha_{2}$ are arbitrary constants.
Naturally, the limit $\varepsilon\rightarrow0$ applied to \eqref{eq:invtensor2}
leads to the non-vanishing components of the invariant tensor for
the Maxwell algebra \eqref{it1}.

From the gauge potential \eqref{one-form2} and the non-vanishing
components of the invariant tensor \eqref{eq:invtensor2}, the Chern-Simons
action \eqref{CSaction} reduces to the torsional Maxwell CS
action 
\begin{equation}
\begin{split}I_{\mathfrak{tor}-\mathfrak{max}}=\frac{1}{16\pi G}\int_{\mathcal{M}} & \alpha_{0}\left(\omega^{a}d\omega_{a}+\frac{1}{3}\epsilon^{abc}\omega_{a}\omega_{b}\omega_{c}\right)+\alpha_{1}\left(2R_{a}e^{a}+\frac{\varepsilon^{2}}{3}\epsilon^{abc}e_{a}e_{b}e_{c}+\varepsilon T^{a}e_{a}\right)\\
 & +\alpha_{2}\left(T^{a}e_{a}+2R^{a}\sigma_{a}+\frac{\varepsilon}{3}\epsilon^{abc}e_{a}e_{b}e_{c}\right)\,,
\end{split}
\label{I-MaxTor}
\end{equation}
up to a surface term. One can see that the CS action is proportional
to three independent sectors, each one with its respective coupling
constant. In particular, the first two terms along $\alpha_{0}$ and
$\alpha_{1}$ correspond to those of the Riemann-Cartan gravity, namely
gravity with non-vanishing torsion. The piece along the $\alpha_{2}$
contains a torsional term, a cosmological term and a term involving
the gravitational Maxwell field $\sigma_{a}$.

Varying the action \eqref{I-MaxTor} with respect to the fundamental
fields yields the field equations of Maxwell CS with torsion,
\begin{eqnarray}
\delta e^{a} & : & \qquad0=\alpha_{1}\left(R_{a}+\varepsilon\hat{T}_{a}\right)+\alpha_{2}\hat{T}_{a}\,,\nonumber \\
\delta\omega^{a} & : & \qquad0=\alpha_{0}R_{a}+\alpha_{1}\hat{T}_{a}+\alpha_{2}F_{a}\,,\label{Eqs.M-MaxTor}\\
\delta\sigma^{a} & : & \qquad0=\alpha_{2}R_{a}\,.\nonumber 
\end{eqnarray}
Imposing the conditions $\alpha_{2}\neq0$ and $\alpha_{2}\neq-\varepsilon\alpha_{1}$
to avoid degeneracy, the previous equations reduce to the vanishing
of the curvature two-forms \eqref{eq:Curvatures-MaxTor}, 
\begin{equation}
\begin{split}R^{a} & =0\,,\\
T^{a}+\frac{\varepsilon}{2}\epsilon^{abc}e_{b}e_{c} & =0\,,\\
D\sigma^{a}+\frac{1}{2}\epsilon^{abc}e_{b}e_{c} & =0\,.
\end{split}
\label{eq:EOM}
\end{equation}
As we can see, the field equations are those of Riemann-Cartan gravity
(with zero curvature and constant torsion) plus the equation of motion
involving the gravitational Maxwell field. Then, the geometries described
by the equations \eqref{eq:EOM} are non-Riemannian (with non-vanishing
torsion) and locally flat. As expected, when $\varepsilon\rightarrow0$,
the above equations reduce to the field equations for the Maxwell
CS gravity theory \eqref{eq:Eqs.M-Max}. Furthermore, note that the
spin connection $\omega^{a}$ can be decomposed as 
\begin{equation}
\omega^{a}=\tilde{\omega}^{a}+k^{a}\,,
\end{equation}
where $\tilde{\omega}^{a}$ corresponds to the torsion-free Levi-Civita
connection, and $k^{a}$ is the contorsion. By the second equation
in \eqref{eq:EOM}, the contorsion is fixed as follows 
\begin{eqnarray}
k^{a}=-\frac{\varepsilon}{2}e^{a}\,.
\end{eqnarray}
Then, the equation of motion of the gravitational Maxwell field and
the Riemann-Cartan curvature $R^{a}$ expressed in terms of its Riemannian
part $\tilde{R}^{a}$, can be written as 
\begin{equation}
\tilde{R}^{a}=\frac{\Lambda_{\text{eff}}}{2}\epsilon^{abc}e_{b}e_{c}\quad;\quad\tilde{D}\sigma^{a}+\frac{1}{2}\epsilon^{abc}\left(e_{b}-2\sqrt{-\Lambda_{\text{eff}}}\,\sigma_{b}\right)e_{c}\;,\label{Riem}
\end{equation}
where the effective cosmological constant is 
\begin{equation}
\Lambda_{\text{eff}}=-\frac{\varepsilon^{2}}{4}\,.
\end{equation}
When the effective cosmological constant is negative, $\Lambda_{\text{eff}}=-\frac{1}{\ell^{2}}$,
the metric is given by the anti-de Sitter solution since $\Lambda_{\text{eff}}<0$.
Indeed, in this case, the solution of the first two equations in \eqref{eq:EOM}
is given by the so-called Riemman-Cartan black hole \cite{Garcia:2003nm}.
In particular, for $\varepsilon=-2/\ell$, this solution leads to
the teleparallel black hole \cite{Blagojevic:2003uc,Blagojevic:2006jk}.

Note that each term of the action \eqref{I-MaxTor} is invariant under
the gauge transformation laws of the algebra \eqref{MaxTor-Alg}.
Indeed, considering the following gauge parameter 
\begin{equation}
\Lambda=\varepsilon^{a}P_{a}+\rho^{a}J_{a}+\chi^{a}Z_{a},
\end{equation}
we have that the gauge transformations $\delta A=d\Lambda+[A,\Lambda]$
of the theory are given by 
\begin{equation}
\begin{split} & \delta_{\Lambda}e^{a}=D\varepsilon^{a}-\epsilon^{abc}\rho_{b}e_{c}+\varepsilon\epsilon^{abc}e_{b}\varepsilon_{c},\\
 & \delta_{\Lambda}\omega^{a}=D\rho^{a},\\
 & \delta_{\Lambda}\sigma^{a}=D\chi^{a}+\epsilon^{abc}e_{b}\varepsilon_{c}-\epsilon^{abc}\rho_{b}\sigma_{c}.
\end{split}
\end{equation}

In the next section, we shall demonstrate that three-dimensional Maxwell
gravity with torsion admits black hole solutions. We show that the
aforementioned solution constitutes a generalization of the teleparallel
black hole presented in \cite{Garcia:2003nm,Blagojevic:2003vn,Blagojevic:2003uc,Mielke:2003xx},
whose thermodynamic properties are directly influenced by the presence
of the gravitational Maxwell field $\sigma^{a}$. Prior to this analysis,
we will present a detailed study of the asymptotic structure of the
previously discussed theory of gravity with torsion.

\subsection{Asymptotic symmetries}

The asymptotic conditions for Maxwell CS gravity theory with non-vanishing
torsion were explored in \cite{Adami:2020xkm}. Here we generalized
this analysis to incorporate chemical potentials conjugated to the
conserved charges and for a generic deformation parameter $\varepsilon$.
Thus, along the line of \cite{Coussaert:1995zp} the radial dependence
of the gauge field is eliminated asymptotically for an appropriate
gauge choice, so that
\begin{equation}
A=g^{-1}dg+g^{-1}a_{(\varepsilon)}g\;,\label{gaugetransformation}
\end{equation}
with $a_{(\varepsilon)}=a_{t}(t,\phi)dt+a_{\phi}(t,\phi)d\phi$. In
this case, the form of $a_{\phi}$ is chosen to be given by \eqref{a_phi-Max}
\cite{Adami:2020xkm}. It is found then that the asymptotic form of
angular component $a_{\phi}$ is left invariant for the Lie-algebra-valued
parameter $\lambda_{(\varepsilon)}=\lambda_{(\varepsilon)}[y,f,h]$
given by 
\begin{equation}
\lambda_{(\varepsilon)}=\lambda_{(0)}+\frac{\varepsilon}{2}\mathcal{N}fP_{0}\;,\label{eq:lambda-MaxTor}
\end{equation}
and providing that the functions $\mathcal{M},\mathcal{N}$ and $\mathcal{F}$
now transform according to: 
\begin{align}
\delta_{(\varepsilon)}{\cal M} & =\delta_{(0)}{\cal M}\;,\nonumber \\
\delta_{(\varepsilon)}{\cal N} & =\delta_{(0)}{\cal N}+\varepsilon\left({\cal N}^{\prime}f+2\mathcal{N}f^{\prime}\right)\;,\label{TransfLaw-MNF-Maxtor}\\
\delta_{(\varepsilon)}{\cal F} & =\delta_{(0)}{\cal F}\;,\nonumber 
\end{align}
where $\lambda_{(0)}$ and $\delta_{(0)}(\cdot)$ are specified by
\eqref{eq:lambda-Max} and \eqref{TransfLaw-MNF-Max}, respectively.
Thus, to preserve the asymptotic symmetries under evolution in time,
the asymptotic form of the Lagrange multiplier $a_{t}$ must to be
of the form
\begin{equation}
a_{t}=\lambda_{(\varepsilon)}[\mu,\xi,\nu]\,,\label{at-MaxTor}
\end{equation}
where the arbitrary functions $\mu(t,\phi)$, $\xi(t,\phi)$ and $\nu(t,\phi)$
are identified with the chemical potentials conjugated to the charges
and which are assumed to be fixed at the boundary \cite{Henneaux:2013dra,Bunster:2014mua}.
The asymptotic behavior of $a_{t}$ is then preserved under the asymptotic
symmetries considering that the field equations:
\begin{align}
\dot{\mathcal{M}} & =2\mathcal{M}\mu^{\prime}+\mathcal{M}^{\prime}\mu-2\mu^{\prime\prime\prime}\,,\nonumber \\
\dot{\mathcal{N}} & =2\mathcal{M}\xi^{\prime}+\mathcal{M}^{\prime}\xi+2\mathcal{N}\mu^{\prime}+\mathcal{N}^{\prime}\mu-2\xi^{\prime\prime\prime}+\varepsilon\left(\mathcal{N}^{\prime}\xi+2\mathcal{N}\xi^{\prime}\right)\,,\label{eq:EqsMotion-MNF-MaxTor}\\
\dot{\mathcal{F}} & =2\mathcal{M}\nu^{\prime}+\mathcal{M}^{\prime}\nu+2\mathcal{N}\xi^{\prime}+\mathcal{N}^{\prime}\xi+2\mathcal{F}\mu^{\prime}+\mathcal{F}^{\prime}\mu-2\nu^{\prime\prime\prime}\,,\nonumber 
\end{align}
are satisfied in the asymptotic region and the parameters $y,f$,
and $h$ fulfill the following conditions: 
\begin{align}
\dot{y} & =y\mu^{\prime}-\mu y^{\prime}\quad;\quad\dot{\xi}=f\mu^{\prime}-\mu f^{\prime}+y\xi^{\prime}-\xi y^{\prime}+\varepsilon\left(f\xi^{\prime}-\xi f^{\prime}\right)\;,\nonumber \\
\dot{\nu} & =h\mu^{\prime}-h^{\prime}\mu+f\xi^{\prime}-f^{\prime}\xi+y\nu^{\prime}-y^{\prime}\nu\;.
\end{align}
The asymptotic symmetry generators are then obtained through the Regge-Teitelboim
method \cite{REGGE1974286} through the expression in \eqref{eq:deltaQ}
considering the invariant tensor in \eqref{eq:invtensor2}, the angular
component of gauge field $a_{\phi}$ in \eqref{a_phi-Max}, and the
algebra-valued parameter defined in \eqref{eq:lambda-MaxTor}. Thus,
one finds that
\begin{equation}
\delta Q[y,f,h]=-\int d\phi\left(y\delta\mathcal{J}+f\delta\mathcal{P}+h\delta\mathcal{Z}\right),\label{charge-variation-01}
\end{equation}
where the canonical generators of the asymptotic symmetry are found
to be determined by 
\begin{equation}
\mathcal{J}=\frac{k}{4\pi}\,\left(\alpha_{2}{\cal F}+\alpha_{1}{\cal N}+\alpha_{0}{\cal M}\right)\quad;\quad\mathcal{P}=\frac{k}{4\pi}\,\left[\left(\varepsilon\alpha_{1}+\alpha_{2}\right){\cal N}+\alpha_{1}{\cal M}\right]\quad;\quad\mathcal{Z}=\frac{k}{4\pi}\,\alpha_{2}{\cal M}\,.\label{JPZtorMax}
\end{equation}
The corresponding transformations laws of these latter canonical generators
can be obtained using \eqref{TransfLaw-MNF-Maxtor}, leading to
\begin{align}
\delta\mathcal{J} & ={\cal J}^{\prime}y+2{\cal J}y^{\prime}+{\cal P}^{\prime}f+2{\cal P}f^{\prime}+{\cal Z}^{\prime}h+2{\cal Z}h^{\prime}-\frac{\alpha_{0}k}{2\pi}y{}^{\prime\prime\prime}-\frac{\alpha_{1}k}{2\pi}f{}^{\prime\prime\prime}-\frac{\alpha_{2}k}{2\pi}h{}^{\prime\prime\prime}\,,\nonumber \\
\delta\mathcal{P} & ={\cal P}^{\prime}y+2{\cal P}y^{\prime}+({\cal Z}^{\prime}+\varepsilon{\cal P}^{\prime})f+2({\cal Z+\varepsilon{\cal P}})f^{\prime}-\frac{\alpha_{1}k}{2\pi}y{}^{\prime\prime\prime}-\frac{(\alpha_{2}+\varepsilon\alpha_{1})k}{2\pi}f{}^{\prime\prime\prime}\,,\\
\delta\mathcal{Z} & ={\cal Z}^{\prime}y+2{\cal Z}y^{\prime}-\frac{\alpha_{2}k}{2\pi}y{}^{\prime\prime\prime}\,.\nonumber 
\end{align}
Under the assumption that the functions $y$, $f$, and $h$ exhibit
no field dependence, the integration of the variation \eqref{charge-variation-01}
becomes trivial, resulting in 
\begin{equation}
Q[y,f,h]=-\int d\phi\left(y\mathcal{J}+f\mathcal{P}+h\mathcal{Z}\right).\label{Q-MaxTor}
\end{equation}
As in the case without torsion, expanding in Fourier modes according
to $X=\frac{1}{2\pi}\sum_{m}X_{m}e^{im\theta}$, one finds that ${\cal J}$,
${\cal P}$ and ${\cal Z}$ obey, in terms of the Poisson bracket,
the following asymptotic symmetry algebra 
\begin{align}
i\left\{ \mathcal{J}_{m},\mathcal{J}_{n}\right\}  & =\left(m-n\right)\mathcal{J}_{m+n}+c_{\mathcal{J}}\,m(m^{2}-1)\delta_{m+n,0}\,,\nonumber \\
i\left\{ \mathcal{J}_{m},\mathcal{P}_{n}\right\}  & =\left(m-n\right)\mathcal{P}_{m+n}+c_{\mathcal{P}}\,m(m^{2}-1)\delta_{m+n,0}\,,\nonumber \\
i\left\{ \mathcal{P}_{m},\mathcal{P}_{n}\right\}  & =\left(m-n\right)(\mathcal{Z}_{m+n}+\varepsilon\mathcal{P}_{m+n})+\left(c_{\mathcal{Z}}+\varepsilon c_{\mathcal{P}}\right)\,m(m^{2}-1)\delta_{m+n,0}\,,\label{MaxTorAlgModes}\\
i\left\{ \mathcal{J}_{m},\mathcal{Z}_{n}\right\}  & =\left(m-n\right)\mathcal{Z}_{m+n}+c_{\mathcal{Z}}\,m(m^{2}-1)\delta_{m+n,0}\,,\nonumber 
\end{align}
where classical central charges $c_{\mathcal{J}},c_{\mathcal{P}}$
and $c_{\mathcal{Z}}$ are defined in \eqref{centralc}. The infinite-dimensional
algebra \eqref{MaxTorAlgModes} can be seen as an infinite-dimensional
enhancement of the deformed Maxwell algebra \eqref{MaxTor-Alg} in
which both the abelian generators $\mathcal{Z}_{n}$ and the generators
$\mathcal{P}_{n}$ have conformal weight $2$. In particular, in the
limit $\varepsilon\rightarrow0$ we recover the asymptotic symmetry
algebra in \eqref{eq:MaxAlgModes} of the three-dimensional Maxwell
CS gravity theory reviewed in Section \eqref{subsec:AsympCond-Max}.
Furthermore, as it was shown in \cite{Adami:2020xkm} the algebra
\eqref{MaxTorAlgModes} is isomorphic to the $\mathfrak{bms}_{3}\oplus\mathfrak{vir}$
algebra. 

In this context, it is important to highlight those gauge field configurations
with $\mathcal{M}$, $\mathcal{N}$, $\mathcal{F}$, and also the
chemical potentials $\xi$, $\mu$, $\nu$ fixed to constants, since
these fulfill the field equations in \eqref{eq:EqsMotion-MNF-MaxTor}.
The gauge field for this class of configurations $a_{(\varepsilon)}=a_{t}dt+a_{\phi}d\phi$,
explicitly reads 
\begin{align}
a_{(\varepsilon)}& =\left(J_{1}+\frac{1}{2}{\cal \,M}\,J_{0}+\frac{1}{2}{\cal N}\,P_{0}+\frac{1}{2}\mathcal{F}\,Z_{0}\right)d\phi+\left[\mu J_{1}+\frac{\mu}{2}\mathcal{M}J_{0}\right.\nonumber \\
 & \left.+\frac{1}{2}\left(\xi\mathcal{M}+\mu\mathcal{N}+\frac{\varepsilon}{2}\mathcal{N}\xi\right)P_{0}+\xi P_{1}+\frac{1}{2}\left(\nu\mathcal{M}+\mu\mathcal{F}+\xi\mathcal{N}\right)Z_{0}+\nu Z_{1}\right]dt\,.\label{a-BTZ-MaxTor}
\end{align}
In appendix \ref{sec:MetricF-MaxTor} by restoring the radial dependence
of the gauge field \eqref{a-BTZ-MaxTor} is shown that for  $\varepsilon=-2/\ell$, where $\ell$ is the AdS$_{3}$ radius, this stationary
solution corresponds to the BTZ black hole \cite{Banados:1992gq}
together to a spin-2 gauge field, named here as the gravitational
Maxwell field. Generically, as follows from \eqref{Q-MaxTor} this
solution not only possesses mass and angular momentum, parametrized
by the spin-2 charges ${\cal P}$, ${\cal J}$, respectively, but
also possesses an extra global charge of spin 2, determined by ${\cal Z}$.
In what follows, we will carry out the calculation and analysis of
the entropy associated with the black hole solution \eqref{a-BTZ-MaxTor}
for Maxwell CS gravity with torsion, as well as examine the corresponding
first law of thermodynamics.

\subsection{Thermodynamics}

Let us move now to the study of the thermodynamics of the solution
\eqref{a-BTZ-MaxTor} of the Maxwell CS theory with torsion.
As in the Maxwell case, the entropy can be determined from the expression
\eqref{Entropy-CS}. Hence, plugging \eqref{a-BTZ-MaxTor} into the
expression for the entropy \eqref{Entropy-CS}, and using the non-vanishing
components of the invariant tensor \eqref{eq:invtensor2}, one finds
\begin{equation}
S=k\left[\alpha_{0}\mu\mathcal{M}+\alpha_{1}\left(\mu\mathcal{N}+\xi\mathcal{M}+\varepsilon\xi\mathcal{N}\right)+\alpha_{2}\left(\nu\mathcal{M}+\xi\mathcal{N}+\mu\mathcal{F}\right)\right]_{\text{on-shell}}\,,\label{S-MaxTor-OffS}
\end{equation}
where it is assumed that the regularity conditions are imposed. 

As in the Maxwell case, and along the lines of \cite{Matulich:2014hea},
we have to find an permissible gauge transformation allowing to gauge
away some temporal components of the gauge field so that it takes
values only in the Lorentz subalgebras of the gauge group \eqref{MaxTor-Alg}.
In this case, for simplicity and exploiting the fact that the deformed
Maxwell algebra can be expressed as the direct sum of the Poincaré
and Lorentz algebras (see \eqref{MaxTor-NewB-LTS}), we consider the
following group element 
\begin{equation}
g=e^{\rho T_{2}}\,,
\end{equation}
so that the angular and temporal components of \eqref{a-BTZ-MaxTor},
take the form, 
\begin{align}
a_{\phi} & =L_{1}+S_{1}+\rho T_{1}+\frac{1}{2}\mathcal{M}L_{0}+\frac{1}{2}\left(\varepsilon\mathcal{F}-\rho\mathcal{M}-\mathcal{N}\right)T_{0}+\frac{1}{2}\left(\mathcal{M}+\varepsilon\mathcal{N}\right)S_{0}\,,\nonumber \\
a_{t} & =\mu L_{1}+\frac{1}{2}\mu\mathcal{M}L_{0}+(\mu+\varepsilon\xi)S_{1}+\frac{1}{2}(\mathcal{M}+\varepsilon\mathcal{N})(\mu+\varepsilon\xi)S_{0}+(\rho\mu-\xi+\varepsilon\nu)T_{1}\nonumber \\
 & +\left(\varepsilon\mu\mathcal{F}-\mu\mathcal{N}-\mathcal{M}(\rho\mu+\xi-\varepsilon\nu)\right)T_{0}\,.
\end{align}
It is thus found that the term along $T_{1}$ in $a_{t}$ vanishes
when $\rho$ is given by 
\begin{equation}
\rho=\frac{\xi-\varepsilon\nu}{\mu}\,,
\end{equation}
and that the term along $T_{0}$ is eliminated by fixing the Lagrange
multiplier $\nu$ according to 
\begin{equation}
\nu=\frac{\xi}{\varepsilon}-\frac{(\varepsilon\mathcal{F}-\mathcal{N})}{2\varepsilon\mathcal{M}}\mu\,,
\end{equation}
so that the temporal component $a_{t}$ reduces to 
\begin{equation}
a_{t}=\mu\left(L_{1}+\frac{1}{2}\mathcal{M}L_{0}\right)+\left(\mu+\varepsilon\xi\right)\left[S_{1}+\frac{1}{2}\left(\mathcal{M}+\varepsilon\mathcal{N}\right)S_{0}\right]\;.
\end{equation}
Thus, imposing that the holonomy of the $a_{\tau}$ around the thermal
circle is trivial, which in this case is written as 
\begin{equation}
H=e^{a_{\tau}}\Big|_{\text{on-shell}}=\left(\begin{array}{cc}
(-1)^{n}\mathds{1}_{2\times2} & 0\\
0 & (-1)^{m}\mathds{1}_{2\times2}
\end{array}\right)\,,
\end{equation}
the chemical potentials $\xi$, $\mu$, $\nu$ become fixed as specific
functions of $\mathcal{M},\mathcal{N}$ and $\mathcal{F}$, obtaining
the following results: 
\begin{align}
\mu & =-\frac{2\pi n}{\sqrt{\mathcal{M}}}\,,\nonumber \\
\xi & =\frac{2\pi}{\varepsilon}\left(\frac{n}{\sqrt{\mathcal{M}}}-\frac{m}{\sqrt{{\cal M}+\mathcal{\varepsilon\mathcal{N}}}}\right)\,,\label{MaxTor-RegCond-MNF}\\
\nu & =\frac{2\pi n}{\sqrt{\mathcal{M}}}\left[\frac{1}{\varepsilon^{2}}+\frac{1}{2\mathcal{M}}\left(\mathcal{F}-\frac{\mathcal{N}}{\varepsilon}\right)\right]-\frac{2\pi m}{\varepsilon^{2}\sqrt{{\cal M}+\mathcal{\varepsilon\mathcal{N}}}}\,,\nonumber
\end{align}
where $n,m$ are integers. Note that in the present case, as we are
dealing with a black hole solution, the orientation of the solid torus
is reversed relative to that of the cosmological solution analyzed
in the section \eqref{sec:Max-Termo}. Indeed, it is shown in
appendix \eqref{sec:MetricF-MaxTor} that for ${\cal M}>0$ and ${\mathcal{M}+\varepsilon\mathcal{N}>0}$ the spacetime develops an inner and an event horizon. Furthermore,  for $n=-m=1$, the regularity conditions in \eqref{MaxTor-RegCond-MNF} are in complete agreement
with those found in the metric formalism in \eqref{RegCond-MaxTor-MetricF}
that stem from requiring the absence of conical singularities of the
Euclidean fields at the event horizon. This means that in the present
case $\xi=1/T_{\text{BH}}$, with $T_{\text{BH}}$ the Hawking temperature.

Thus, substituting \eqref{MaxTor-RegCond-MNF} into the entropy \eqref{S-MaxTor-OffS},
one obtains 
\begin{align}
S_{\text{BH}} & =2\pi k\sqrt{\mathcal{M}}\left\{ \left(\frac{\alpha_{1}\varepsilon+\alpha_{2}}{\varepsilon^{2}}\right)\left(n+m\sqrt{1+\frac{\varepsilon\mathcal{N}}{\mathcal{M}}}\right)-n\left[\alpha_{0}+\frac{\alpha_{2}}{2\varepsilon}\left(\frac{\varepsilon\mathcal{F}-\mathcal{N}}{\mathcal{M}}\right)\right]\right\} \;,\label{Entropy-MaxTor}
\end{align}

It can be readily checked that, upon imposing the condition $m=n=-n_{\mathfrak{m}}$,
the expression reduces to \eqref{Entropy-Max} in the limit $\varepsilon\rightarrow0$.
In terms of the global charges $\mathcal{P},\mathcal{J}$ y $\mathcal{Z}$,
the entropy finally reads 
\begin{equation}
S_{\text{BH}}=\frac{4\pi\sqrt{\pi k}\sqrt{\alpha_{2}}}{\varepsilon}\left[\frac{n}{2}\left(\frac{\mathcal{P}-\varepsilon\mathcal{J}}{\sqrt{\mathcal{Z}}}\right)+n\left(\frac{1}{\varepsilon}+\frac{\alpha_{1}-\varepsilon\alpha_{0}}{2\alpha_{2}}\right)\sqrt{\mathcal{Z}}-\frac{m}{\varepsilon}\sqrt{\frac{\alpha_{2}+\varepsilon\alpha_{1}}{\alpha_{2}}}\sqrt{\mathcal{Z}+\varepsilon\mathcal{P}}\right]\,.\label{entropybhfinal}
\end{equation}
The entropy $S_{\text{BH}}$ is real for conditions $\alpha_{2}>0$,
$\alpha_{2}+\varepsilon\alpha_{1}>0$, $\mathcal{Z}>0$ and $\mathcal{Z}+\varepsilon\mathcal{P}>0$.
Moreover, for $\varepsilon<0$, we can see that this quantity is positive
for the additional conditions: 
\begin{itemize}
\item When $m=n$, for $n>0$, $\mathcal{J}>\mathcal{J}_{c}$ and for $n<0$,
$\mathcal{J}<\mathcal{J}_{c}$, with 
 
\end{itemize}
\begin{equation}
\mathcal{J}_{c}=\frac{\mathcal{P}}{\varepsilon}-2\sqrt{\frac{\mathcal{Z}(\text{\ensuremath{\alpha_{1}}}\varepsilon+\text{\ensuremath{\alpha_{2}}})(\varepsilon\mathcal{P}+\mathcal{Z})}{\text{\ensuremath{\alpha_{2}}}\varepsilon^{4}}}+\frac{\mathcal{Z}(\varepsilon(\text{\ensuremath{\alpha_{1}}}-\text{\ensuremath{\alpha_{0}}}\varepsilon)+2\text{\ensuremath{\alpha_{2}}})}{\text{\ensuremath{\alpha_{2}}}\varepsilon^{2}}\,.
\end{equation}

\begin{itemize}
\item When $m=-n$, for $n>0$, $\mathcal{J}>\mathcal{J}_{c}$ and for $n<0$,
$\mathcal{J}<\mathcal{J}_{c}$, with 
 
\end{itemize}
\begin{equation}
\mathcal{J}_{c}=\frac{\mathcal{P}}{\varepsilon}+2\sqrt{\frac{\mathcal{Z}(\text{\ensuremath{\alpha_{1}}}\varepsilon+\text{\ensuremath{\alpha_{2}}})(\varepsilon\mathcal{P}+\mathcal{Z})}{\text{\ensuremath{\alpha_{2}}}\varepsilon^{4}}}+\frac{\mathcal{Z}(\varepsilon(\text{\ensuremath{\alpha_{1}}}-\text{\ensuremath{\alpha_{0}}}\varepsilon)+2\text{\ensuremath{\alpha_{2}}})}{\text{\ensuremath{\alpha_{2}}}\varepsilon^{2}}\,.
\end{equation}
As in the previous analysis, the temperature and chemical potentials
within the microcanonical ensemble are determined using the thermodynamic
relations, 
\begin{align}
\beta_{\text{BH}}= & \left.\left(\frac{\partial S_{\text{BH}}}{\partial\mathbb{M}}\right)\right|_{\mathbb{J},\mathbb{W}}=T_{\text{BH}}^{-1}=\frac{\sqrt{\pi k}}{\varepsilon}\left(\frac{n\sqrt{\alpha_{2}}}{\sqrt{\mathcal{Z}}}-\frac{m\sqrt{\alpha_{2}+\varepsilon\alpha_{1}}}{\sqrt{\varepsilon\mathcal{P}+\mathcal{Z}}}\right)\,,\\
\Omega_{\text{BH}}= & -\beta_{\text{BH}}^{-1}\left.\left(\frac{\partial S_{\text{BH}}}{\partial\mathbb{J}}\right)\right|_{\mathbb{M},\mathbb{W}}=\frac{n\varepsilon\sqrt{\alpha_{2}}\sqrt{\varepsilon\mathcal{P}+\mathcal{Z}}}{m\sqrt{\mathcal{Z}}\sqrt{\alpha_{1}\varepsilon+\alpha_{2}}-n\sqrt{\alpha_{2}}\sqrt{\varepsilon\mathcal{P}+\mathcal{Z}}}\,,\\
\Phi_{\text{BH}}= & -\beta_{\text{BH}}^{-1}\left.\left(\frac{\partial S_{\text{BH}}}{\partial\mathbb{W}}\right)\right|_{\mathbb{M},\mathbb{J}}=\frac{n\left[\alpha_{2}\left(\varepsilon\mathcal{J}-\mathcal{P}\right)+\mathcal{Z}\left(\alpha_{1}-\varepsilon\alpha_{0}\right)\right]\sqrt{\varepsilon\mathcal{P}+\mathcal{Z}}}{2\sqrt{\alpha_{2}}\mathcal{Z}\left(m\sqrt{\mathcal{Z}}\sqrt{\alpha_{1}\varepsilon+\alpha_{2}}-n\sqrt{\alpha_{2}}\sqrt{\varepsilon\mathcal{P}+\mathcal{Z}}\right)}-\frac{1}{\varepsilon}\,,
\end{align}
where $\mathbb{M}=2\pi\mathcal{P}$, $\mathbb{J}=-2\pi\mathcal{J}$,
and $\mathbb{W}=2\pi\mathcal{Z}$. These thermodynamical quantities
satisfy the first law of thermodynamics. 
\begin{equation}
\delta S_{\text{BH}}=\beta_{\text{BH}}(\delta\mathbb{M}-\Omega_{\text{BH}}\delta\mathbb{J}-\Phi_{\text{BH}}\delta\mathbb{W})\,,
\end{equation}
such that the chemical potentials $\beta_{\text{BH}}=\xi$, $\Omega_{\text{BH}}=\mu/\xi$,
and $\Phi_{\text{BH}}=-\nu/\xi$ correspond to the conjugated to the mass ($\mathbb{M}$),
angular momentum  ($\mathbb{J}$) and the additional spin-2 charge  ($\mathbb{W}$), respectively. It is worth pointing out that the mass of the black hole turns out to be  bounded from below when
\begin{equation}
\alpha_{1}\left(r_{+}-r_{-}\right)^{2}+2\alpha_{2}\text{sgn}({\cal N})r_{+}r_{-}>0\;,
\end{equation}
where $r_{+}$ and $r_{-}$ correspond to  the outer and inner horizons, respectively, given in \eqref{rmasmen}. 

To recover the results of the Maxwell theory without torsion in the
limit $\varepsilon\rightarrow0$, it is necessary to impose the condition
$m=n$. As a remark, in the present case we have $\xi=1/T_{\text{BH}}$,
whereas in the cosmological solution of Maxwell CS, the corresponding
relation is $\xi=-1/T_{C}$. As we have mentioned, with the choice
$m=n=-n_{\mathfrak{m}}$, it is straightforward to show that the limit
$\varepsilon\rightarrow0$ applied to the black hole entropy \eqref{entropybhfinal}
leads to the entropy of the cosmological solution of the Maxwell CS
theory \eqref{S-MaxJPZ2}.

It must be emphasized that we can express the black hole entropy \eqref{Entropy-MaxTor} in terms of the outer and inner horizons of the black hole given in \eqref{rmasmen} as follows 
\begin{itemize}
\item $m=n$ 
\begin{equation}
S_{\text{inn}}=2n\pi k\left(\frac{\alpha_{0}}{\ell}\left(r_{+}-r_{-}\right)-\alpha_{1}r_{-}-\frac{\alpha_{2}(r_{-}^{2}-\mathcal{F})\ell}{2(r_{+}-r_{-})}\right)\,,
\label{S-in-MaxTor}
\end{equation}
\item $m=-n$ 
\begin{equation}
S_{\text{out}}=2n\pi k\left(\frac{\alpha_{0}}{\ell}\left(r_{+}-r_{-}\right)+\alpha_{1}r_{+}-\frac{\alpha_{2}(r_{+}^{2}-\mathcal{F})\ell}{2(r_{+}-r_{-})}\right)\,,
\label{S-out-MaxTor}
\end{equation}
\end{itemize}
where we have fixed $\varepsilon=-2/\ell$. We conclude that $S_{\text{out}}$ corresponds to a Maxwell extension
(along $\alpha_{2}$) of the entropy of the teleparallel black hole
extensively analyzed in \cite{Garcia:2003nm,Blagojevic:2006jk,Blagojevic:2006nf,Blagojevic:2006hh},
and recently studied in \cite{Geiller:2020edh}. The first law in
this case takes the form 
\begin{equation}
\delta\mathbb{M}=T_{\text{BH}}\delta S_{\text{out}}+\Omega_{H}\delta\mathbb{J}+\Phi\delta\mathbb{W}
\end{equation}
where the extensive variables can be expressed as follows 
\begin{equation}
T_{\text{BH}}=\frac{r_{+}^{2}-r_{-}^{2}}{2\pi n\ell^{2}r_{+}}\,,\qquad\Omega_{H}=\frac{1}{\ell}+\frac{r_{-}}{\ell r_{+}}\,,\qquad\Phi=\frac{\left(r_{+}^{2}+(1+r_{-}/r_{+})\mathcal{F}-3r_{+}r_{-}\right)\ell}{2(r_{+}-r_{-})^{2}}\,.
\end{equation}

As we have already mentioned, it is not possible to take the flat
limit $\ell\rightarrow\infty$ in $S_{\text{out}}$ since $r_{+}(M,J,\ell)$
is pushed out to infinity and consequently does not have a well-defined
limit. We can consider instead the thermodynamics of the inner horizon
of the Maxwell BTZ-like black hole with torsion as follows, 
\begin{equation}
T_{\text{inn}}=\frac{r_{+}^{2}-r_{-}^{2}}{2\pi n\ell^{2}r_{-}}\,,\qquad\Omega_{\text{inn}}=\frac{1}{\ell}+\frac{r_{+}}{\ell r_{-}}\,,\qquad\Phi_{\text{inn}}=\frac{\left(r_{-}^{2}+(1+r_{+}/r_{-})\mathcal{F}-3r_{+}r_{-}\right)\ell}{2(r_{+}-r_{-})^{2}}\,,
\end{equation}
with a first law of the form 
\begin{equation}
\delta\mathbb{M}={-}T_{\text{inn}}\delta S_{\text{inn}}+\Omega_{\text{inn}}\delta\mathbb{J}+\Phi_{\text{inn}}\delta\mathbb{W}\,.
\end{equation}
The thermodynamics of the inner (Cauchy) horizon of black holes has
been studied in \cite{Castro:2012av,Detournay:2012ug}. As happens
in the standard BTZ black hole, the horizon of the cosmological solution
in the Maxwell case, corresponds to the limit of the inner horizon
of the Maxwell BTZ-like black holes with torsion. Furthermore, as
we have mentioned before all thermodynamic quantities of the Maxwell
cosmological solutions are recovered in the flat limit $\ell\rightarrow\infty$
of the quantities of the inner horizon of the Maxwell torsional BTZ-like
black hole (when $m=n$).

It is also important to point out that in the metric formalism ($n=-m=1$)
the entropy in \eqref{Entropy-MaxTor} and \eqref{S-in-MaxTor}-\eqref{S-out-MaxTor}  can be also fully expressed
as the sum of the horizon area element and its spin-2 analogues \cite{Perez:2013xi,Grumiller:2016kcp}
leading to an extended version of the standard Bekenstein-Hawking
entropy formula given by 
\begin{equation}
S_{\text{BH}}=k\left(\alpha_{0}{\cal A}_{\text{CS}}+\alpha_{1}{\cal A}_{\text{metric}}+\alpha_{2}{\cal A}_{\text{Max}}\right),
\end{equation}
where ${\cal A}_{\text{CS}}$, ${\cal A}_{\text{Max}}$ and ${\cal A}_{\text{metric}}$
are related to the values of the purely angular components of the spin-2 fields at horizons: 
\begin{align}
\left(\frac{{\cal A}_{\text{metric}}}{2\pi}\right)^{2} & =e_{a\phi}e_{\;\phi}^{a}|_{r_{\pm}}=\frac{1}{\varepsilon^{2}}\left(\sqrt{\mathcal{M}}\pm\sqrt{\mathcal{M}+\varepsilon\mathcal{N}}\right)^{2}=r_{\pm}^{2}\;,\nonumber \\
\left(\frac{{\cal A}_{\text{Max}}}{2\pi}\right)^{2} & =\sigma_{a\phi}\sigma_{\;\phi}^{a}|_{r_{\pm}}=\frac{\varepsilon^{2}}{4{\cal M}}\left[\mathcal{F}-\frac{1}{\varepsilon^{2}}\left(\sqrt{\mathcal{M}}\pm\sqrt{\mathcal{M}+\varepsilon\mathcal{N}}\right)^{2}\right]^{2}=\frac{(\mathcal{F}-r_{\pm}^{2})^{2}}{\varepsilon^{2}(r_{+}-r_{-})^{2}}\;,\label{Area-Analog-MaxTor}\\
\left(\frac{{\cal A}_{\text{CS}}}{2\pi}\right)^{2} & =\omega_{a\phi}\omega_{\;\phi}^{a}|_{r_{\pm}}={\cal M}=\frac{1}{4}\varepsilon^{2}\left(r_{+}+\text{sgn}\,({\cal N}) r_{-}\right)^{2}\;,\nonumber 
\end{align}
where these particular contributions in \eqref{Area-Analog-MaxTor}
are defined in terms of the pullback of the metric and the spin-2
fields at the spacelike section of the horizon defined in \eqref{Area-Analog}
(see Refs. \cite{Perez:2013xi,Grumiller:2016kcp}).

\section{Discussion}

\label{concl}

In this work, we performed a comprehensive analysis of the thermodynamics
of three-dimensional gravity solutions based on Maxwell symmetry,
considering both cases with and without torsion. As mentioned in the
Introduction, despite the algebraic richness and theoretical relevance
of Maxwell CS gravity, the thermodynamic behavior of its solutions
had not been thoroughly investigated. In this article, we contributed
to closing this gap by examining the entropy and establishing the
thermodynamic consistency of asymptotically flat cosmological solutions
and black holes in three-dimensional Maxwell Chern-Simons.

We first examined the behavior of the dynamical fields in the asymptotic
region, which was essential for computing the thermodynamic properties
of the solutions. In both scenarios, we considered asymptotic conditions
incorporating chemical potentials conjugate to the conserved charges.
In the case without torsion, we found that the entropy of the cosmological
solution extends the one for General Relativity in the presence of
the CS term, and has a new contribution proportional to the $\alpha_{2}$
constant, showing that the gravitational Maxwell field modifies not
only the conserved charges but also the entropy of the solution.

We then extended the analysis to the thermodynamics of the Maxwell
BTZ-like black hole with torsion, and we found that the entropy generalizes
the one of the torsional BTZ black hole, exhaustively studied in \cite{Garcia:2003nm,Blagojevic:2006jk,Blagojevic:2006nf,Blagojevic:2006hh}.
By evaluating the entropy through the first law, we verified that
the resulting expressions consistently satisfy black hole thermodynamics
in both torsional and torsionless theories. In the appropriate limit,
the charges and entropy of the Maxwell BTZ-like black hole with torsion
reduce smoothly to the results of Maxwell CS without torsion,
which confirms the robustness of our formulation. Taken together,
these results establish a coherent thermodynamic framework for Maxwell
CS gravity theories, clarifying how torsion affects the structure
of conserved charges and the interpretation of physical parameters
in stationary black hole solutions. It would be interesting to extend
our study and the asymptotic symmetry analysis to the non-relativistic
and ultra-relativistic regimes of the Maxwell CS gravity theory \cite{Aviles:2018jzw}.
Of particular interest is the emergence of an infinite-dimensional
extension of the generalized Maxwell algebra as the asymptotic symmetry
algebra of the AdS-Carroll CS gravity \cite{Aviles:2025ygw}. One
could investigate the boundary dynamics of Maxwellian generalizations
of the Galilean/Carroll (super)gravity introduced in \cite{Aviles:2018jzw,Penafiel:2019czp,Concha:2019mxx,Concha:2020ebl,Concha:2020eam,Concha:2021jnn,Caroca:2022byi,Concha:2024tcu},
together with the space of solutions satisfying the set of asymptotic
conditions. Furthermore, it would be natural to consider alternative
boundary conditions in order to explore their impact on both the asymptotic
symmetry algebra and the solution space.

On the other hand, the presence of non-vanishing torsion in a non-relativistic
regime has also been shown to be particularly interesting. In fact,
non-vanishing torsion in such a setting was first encountered in the
context of Lifshitz holography \cite{Christensen:2013lma} and the
Quantum Hall Effect \cite{Geracie:2015dea}. Subsequently, diverse
strategies have been implemented to construct non-relativistic gravity
models with non-vanishing torsion \cite{Bergshoeff:2015ija,Bergshoeff:2017dqq,VandenBleeken:2017rij,Concha:2022you,Concha:2023ejs}.
Thus, it would be worthwhile to explore the boundary dynamics of torsional
non-relativistic gravity models.

It should also be emphasized that the study presented in this article
provides the missing elements necessary to carry out a complete analysis
of the energy bounds and asymptotic Killing spinors in the supersymmetric
extensions of both Maxwell/Hietarinta and AdS--Lorentz gravity theories
\cite{Concha:2018jxx,Matulich:2023xpw,Concha:2023nou}.



\section*{Acknowledgment}

We thank Rodrigo Aros and Ricardo Troncoso for interesting remarks and discussions.
This work was funded by the National Agency for Research and Development
ANID - FONDECYT grants 1250642, 1231133 and 1221624. This work was
supported by the Research project Code DIREG 04/2025 (P.C.) and FGII05/2024
(P.C., J.M. and E.R.) of the Universidad Católica de la Santisima
Concepción (UCSC), Chile. P.C., J.M. and E.R. would also like to thank
the Dirección de Investigación and Vice-rectoría de Investigación
of the Universidad Católica de la Santísima Concepción, Chile, for
their constant support. J.M. has been supported by the MCI, AEI, FEDER
(UE) grants PID2021-125700NB- C21 (“Gravity, Supergravity and Superstrings”
(GRASS)) and IFT Centro de Excelencia Severo Ochoa CEX2020-001007-S.
L.A. has been supported by SIA-ANID grant 85220027.


\appendix

\section{Chemical Potentials in the Metric Formalism (Torsionless Case)}\label{sec:MetricF-Max}

In order to recovered the spacetime metric $g_{\mu\nu}$ as well as
the gravitational Maxwell field $\sigma_{\mu\nu}$ the radial dependence
in the gauge field \eqref{eq:a-Cosmo-Mx} must be introduced through
a permissible gauge transformation. This procedure can be achieved
by choosing the following group element 
\begin{equation}
g_{{\cal C}}=e^{F(r)P_{2}+G(r)Z_{2}}\;,
\end{equation}
where the functions $F(r)$ and $G(r)$ are defined according to 
\begin{align}
F(r) & =\frac{\mathcal{N}}{2\mathcal{M}}\left(1+\sqrt{1-\frac{4{\cal M}}{{\cal N}^{2}}r^{2}}\right)\;;\quad G(r)=F(r)+\frac{1}{2\mathcal{M}}\left(\mathcal{F}-\mathcal{N}-\frac{{\cal N}^{2}}{2{\cal M}}\right)\;,
\end{align}
Thus, the full gauge field 
\begin{equation}
A=\omega^{a}J_{a}+e^{a}P_{a}+\sigma^{a}Z_{a}=g_{{\cal C}}^{-1}ag_{{\cal C}}+g_{{\cal C}}^{-1}dg_{{\cal C}}\,.
\end{equation}
is so that their spacetime components can be suitably written as follows
\begin{equation}
A_{\phi}=a_{\phi}+c_{\phi}(r)\quad;\quad A_{t}=a_{t}+c_{t}(r)\quad;\quad A_{r}=c_{r}(r)\;,
\end{equation}
where $a_{\phi}$ and $a_{t}$ given in \eqref{eq:a-Cosmo-Mx} while
$c_{\phi}(r)$, $c_{\phi}(r)$ and $c_{r}(r)$ given by 
\begin{align}
c_{\phi} & =F(r)\left(P_{1}-\frac{1}{2}\mathcal{M}P_{0}-\frac{1}{2}\mathcal{N}Z_{0}\right)+\left(G(r)+\frac{1}{2}F(r)^{2}\right)\left(Z_{1}-\frac{1}{2}\mathcal{M}Z_{0}\right)+\frac{1}{2}F(r)^{2}\mathcal{M}Z_{0}\;,\nonumber \\
c_{t} & =F(r)\left[\left(P_{1}-\frac{1}{2}\mathcal{M}P_{0}-\frac{1}{2}\mathcal{N}Z_{0}\right)\mu+\left(\xi+\frac{G(r)}{F(r)}\mu\right)\left(Z_{1}-\frac{1}{2}\mathcal{M}Z_{0}\right)\right]\nonumber \\
 & +\frac{1}{2}\mu F(r)^{2}\left(Z_{1}+\frac{1}{2}\mathcal{M}Z_{0}\right)\;,\\
c_{r} & =F(r)^{\prime}P_{2}+G(r)^{\prime}Z_{2}\;.\nonumber 
\end{align}
The spacetime metric defined by 
\begin{equation}
ds^{2}=\eta_{ab}e_{\;\mu}^{a}e_{\;\nu}^{b}dx^{\mu}dx^{\nu}\;,
\end{equation}
is then directly obtained in Schwarzschild-like coordinates, 
\begin{equation}
ds^{2}=-\left(\frac{{\cal N}^{2}}{4r^{2}}-{\cal M}\right)\xi^{2}dt^{2}+\left(\frac{{\cal N}^{2}}{4r^{2}}-{\cal M}\right)^{-1}dr^{2}+r^{2}\left[d\phi+\left(\mu+\frac{{\cal N}\xi}{2r}\right)dt\right]^{2}\;,\label{ds2Max}
\end{equation}
where for ${\cal M}>0$, it possesses a cosmological horizon located
at 
\begin{equation}
r=r_{c}=\frac{\left|{\cal N}\right|}{2\sqrt{{\cal M}}}\;.
\end{equation}
In this frame the nontrivial components of spin connection are given by
\begin{equation}
\omega^{0}=\frac{1}{2}{\cal M}\left(d\phi+\mu dt\right)\;,\qquad\omega^{1}=d\phi+\mu dt\;\label{Max-w-Rc}.
\end{equation}
The Maxwell CS field defined here as $\sigma_{\mu\nu}=\eta_{ab}e_{\;\mu}^{a}\sigma_{\;\nu}^{b}$,
can be suitably expressed as follows 
\begin{equation}
d\sigma^{2}=\eta_{ab}e_{\;\mu}^{a}\sigma_{\;\nu}^{b}dx^{\mu}dx^{\nu}=\left\{ \varrho_{\mu\nu}+\frac{3}{4}\xi\left[r^{2}+\frac{1}{3}\left({\cal F}-\frac{\nu}{\xi}{\cal N}\right)\right]\epsilon_{r\mu\nu}\right\} dx^{\mu}dx^{\nu}\;,\label{eq:dsigma2-Max}
\end{equation}
where its symmetry part $\sigma_{(\mu\nu)}$ reads 
\begin{align}
\varrho_{\mu\nu}dx^{\mu}dx^{\nu} & =\xi\mathcal{M}\left(\nu+\frac{1}{2}\frac{\mathcal{N}}{\mathcal{M}}\xi\right)dt^{2}+\left[\left(1+\frac{1}{4}\frac{\mathcal{N}}{\mathcal{M}}\right)r^{2}+\frac{1}{4}\frac{\mathcal{N}}{\mathcal{M}}\left(\mathcal{F}-\mathcal{N}-\frac{\mathcal{N}^{2}}{2\mathcal{M}}\right)\right]\left(d\phi+\mu dt\right)^{2}\nonumber \\
 & +\frac{1}{2}\left(\xi r^{2}+\xi\mathcal{F}+\nu\mathcal{N}\right)\left(d\phi+\mu dt\right)dt+\left(\frac{{\cal N}^{2}}{4r^{2}}-{\cal M}\right)^{-1}dr^{2}\;.
\end{align}
Note that the chemical potentials $\xi$, $\mu$ and $\nu$ have been
explicitly incorporated in the metric $g_{\mu\nu}$ and the gravitational
CS field $\sigma_{\mu\nu}$, so regularity analysis of the fields
around the cosmological horizon at $r=r_{c}$ can be carried out for
a fixed range of the angular coordinates of the solid torus, where
it is assumed to be $0<\tau\leq1$, and $0<\phi\leq2\pi$. It is follows
that around the cosmological horizon the rotating frame in \eqref{ds2Max}
can be fixing by choosing the Lagrange multiplier $\mu$ according
to 
\begin{equation}
\mu=-\frac{{\cal N}\xi}{2r_{c}}=-\text{sgn}\left({\cal N}\right)\sqrt{\mathcal{M}}\xi\;,\label{eq:RotFrame-mu}
\end{equation}
so that expanding around the cosmological horizon the metric reduces
to Rindler space\footnote{This step can be carried out by redefining the radial coordinate according
to $r=r_{c}\,\text{sech}\left(\frac{2\mathcal{M}}{\mathcal{N}}\rho\right)$.}, that is 
\begin{equation}
ds^{2}\approx-\frac{4\mathcal{M}^{3}\xi^{2}}{\mathcal{N}^{2}}\rho^{2}dt^{2}+d\rho^{2}+\frac{1}{4}\frac{\mathcal{N}^{2}}{\mathcal{M}}d\phi^{2}\;,\label{eq:Metric-Rc}
\end{equation}
from which the temperature can be directly read off and it is given
by 
\begin{equation}
\xi^{2}=\frac{\mathcal{N}^{2}\pi^{2}}{\mathcal{M}^{3}}\;.\label{eq:xi2-Rc}
\end{equation}
The gravitational Maxwell field \eqref{eq:dsigma2-Max} around the
horizon behaves as follows 
\begin{equation}
d\sigma^{2}\approx\left[\varrho_{\mu\nu}+\frac{1}{4}\left(\nu-\left(\mathcal{F}-\frac{3}{4}\frac{\mathcal{N}^{2}}{\mathcal{M}}\right)\frac{\xi}{\mathcal{N}}\right)\mathcal{N}\epsilon_{r\mu\nu}\right]dx^{\mu}dx^{\nu}\;,\label{eq:MaxF-Rc}
\end{equation}
where 
\begin{align}
\varrho_{\mu\nu}dx^{\mu}dx^{\nu} & \approx-\frac{4\mathcal{M}^{3}\xi^{2}}{\mathcal{N}^{2}}\rho^{2}dt^{2}+d\rho^{2}+\frac{1}{2}\mathcal{N}\left[\nu-\left(\mathcal{F}-\frac{3\mathcal{N}^{2}}{4\mathcal{M}}\right)\mathcal{N}^{-1}\xi\right]dtd\phi+\frac{1}{4}\frac{\mathcal{N}}{\mathcal{M}}\left(\mathcal{F}-\frac{\mathcal{N}^{2}}{4\mathcal{M}}\right)d\phi^{2}\;,\label{eq:MaxF-Rc2}
\end{align}
Thus, fixing the rotating frame in the Maxwell CS field \eqref{eq:MaxF-Rc}
at the horizon leads to that the Lagrange multiplier $\nu$ becoming
determined by 
\begin{align}
\nu & =\pi\left(\mathcal{F}-\frac{3}{4}\frac{\mathcal{N}^{2}}{\mathcal{M}}\right)\frac{\xi}{\mathcal{N}}\;.\label{eq:RotFrame-nu}
\end{align}

It is worth pointing out that both the metric and the gravitational
Maxwell field are finite and well-behaved at the horizon. Indeed,
around the horizon both the metric \eqref{eq:Metric-Rc} and the gravitational
Maxwell field \eqref{eq:MaxF-Rc} by virtue of the regularity conditions
\eqref{eq:RotFrame-mu} and \eqref{eq:RotFrame-nu} reduce to Rindler
spacetimes with the same Hawking temperature which becomes fixed to
$\xi=-1/T_{C}$, where the minus sign amounts to the fact that orientation
of the solid torus is reversed compared to that of the black hole;
so that the chemical potential $\xi$ corresponds to the minus branch
of \eqref{eq:xi2-Rc}. This latter feature provides the additional condition  $\varrho_{\phi\phi}\left(r_{c}\right)>0$, which would mean that both fields really share the same topology around the horizon, that is,
\begin{equation}
\varrho_{\phi\phi}\left(r_{c}\right)=\frac{\mathcal{N}}{\mathcal{M}}\left(\mathcal{F}-\frac{\mathcal{N}^{2}}{4\mathcal{M}}\right)>0\;,
\end{equation}
which follows from \eqref{eq:MaxF-Rc2}.

Thus, in the metric formalism the regularity conditions on the fields are found to be given by 
\begin{align}
\xi & =-\frac{\pi\left|\mathcal{N}\right|}{\mathcal{M}^{3/2}}\,, & \mu & =\frac{2\pi}{\mathcal{M}^{1/2}}\text{sgn}\left({\cal N}\right)\,, & \nu & =-\pi\left[\frac{\mathcal{F}}{\mathcal{M}^{3/2}}-\frac{3\mathcal{N}^{2}}{4\mathcal{M}^{5/2}}\right]\text{sgn}\left({\cal N}\right)\,.\label{eq:RegCond-Max-MetricF}
\end{align}
As a close remark one can verifies from \eqref{eq:Metric-Rc}, \eqref{eq:MaxF-Rc2}
and \eqref{Max-w-Rc} that the purely angular components of following
spin-2 fields at the cosmological horizon, 
\begin{align}
g_{\phi\phi}\left(r_{c}\right) & =e_{a\phi}e_{\;\phi}^{a}|_{r_{c}}=\left(\frac{{\cal A}_{\text{metric}}}{2\pi}\right)^{2}=\frac{\mathcal{N}^{2}}{4\mathcal{M}}\;,\nonumber \\
h_{\phi\phi}\left(r_{c}\right) & =\sigma_{a\phi}\sigma_{\;\phi}^{a}|_{r_{c}}=\left(\frac{{\cal A}_{\text{Max}}}{2\pi}\right)^{2}=\left(\frac{\mathcal{F}}{\sqrt{\mathcal{M}}}-\frac{\mathcal{N}^{2}}{4\mathcal{M}^{3/2}}\right)^{2}\;,\label{eq:AngComp-MaxFs}\\
w_{\phi\phi}\left(r_{c}\right) & =\omega_{a\phi}\omega_{\;\phi}^{a}|_{r_{c}}=\left(\frac{{\cal A}_{\text{CS}}}{2\pi}\right)^{2}={\cal M}\;,\nonumber 
\end{align}
appear to be finite and along the lines of \cite{Perez:2013xi,Grumiller:2016kcp}
can be precisely defined in terms of the pullback of the metric and
the additional spin-2 fields at the spacelike section of the horizon
according to
\begin{equation}
{\cal A}=\int_{\partial\Sigma_{+}}\left(X_{\mu\nu}\frac{dx^{\mu}}{d\sigma}\frac{dx^{\nu}}{d\sigma}\right)^{1/2}d\sigma\;,\label{Area-Analog}
\end{equation}
which turns out to be reparametrization invariant. Indeed, it is reassuring to find in the same manner as in \cite{Solodukhin:2005ah } (see e.g. \cite{Bonora:2011gz}) that $\cal{A}_{\text{CS}}$ turns out to be Lorentz invariant in spite the fact that the Lorentz connection enters explicitly.

\section{Chemical Potentials in the Metric Formalism with Torsion}\label{sec:MetricF-MaxTor}

For the Maxwell CS theory with torsion \eqref{I-MaxTor} the corresponding
spacetime metric $g_{\mu\nu}$ as well as the gravitational Maxwell
field $\sigma_{\mu\nu}$ are constructed through the restoring the
radial dependence in the full gauge field. For the solution in \eqref{a-BTZ-MaxTor}
the latter can be carried out by the following group element
\begin{equation}
g_{{\cal B}}=e^{F(r)P_{2}+G(r)Z_{2}}\;,
\end{equation}
with $F(r)$ and $G(r)$ given by 
\begin{align}
F(r) & =\varepsilon^{-1}\text{log}\left[1-\frac{\varepsilon r}{{\cal M}}\left(\frac{\sigma r^{2}-{\cal N}}{2r}-f(r)\right)\right]\;;\quad G(r)=F(r)+\frac{\varepsilon\mathcal{F}-\mathcal{N}}{2\varepsilon\mathcal{M}}\;,
\end{align}
and where the function $f(r)$ reads 
\begin{equation}
f(r)^{2}=\frac{\varepsilon^{2}}{4}r^{2}-\left(\mathcal{M}+\frac{1}{2}\varepsilon\mathcal{N}\right)+\frac{\mathcal{N}^{2}}{4r^{2}}=\left(\frac{\varepsilon r^{2}-\mathcal{N}}{2r}\right)^{2}-\mathcal{M}\;.\label{f(r)-MaxTor}
\end{equation}
Thus, the components of the full gauge field, 
\begin{equation}
A=\omega^{a}J_{a}+e^{a}P_{a}+\sigma^{a}Z_{a}=g_{{\cal B}}^{-1}ag_{{\cal B}}+g_{{\cal B}}^{-1}dg_{{\cal B}}\,,
\end{equation}
are suitably written as follows 
\begin{equation}
A_{\phi}=a_{\phi}+c_{\phi}(r)\quad;\quad A_{t}=a_{t}+c_{t}(r)\quad;\quad A_{r}=c_{r}(r)\;,
\end{equation}
where $a_{\phi}$ and $a_{t}$ given in \eqref{a-BTZ-MaxTor} while
$c_{\phi}(r)$, $c_{\phi}(r)$ and $c_{r}(r)$ read 
\begin{align}
c_{\phi} & =2\sigma^{-2}e^{\varepsilon F(r)/2}\text{sinh}\left(\frac{\varepsilon}{2}F(r)\right)\left[Z_{1}+\varepsilon P_{1}-\frac{1}{2}\left(\mathcal{M}+\varepsilon{\cal N}\right)e^{-\varepsilon F(r)}\left(Z_{0}+\sigma P_{0}\right)\right]\nonumber \\
 & -\varepsilon^{-1}\left(F(r)-\varepsilon G(r)\right)\left[Z_{1}-\frac{1}{2}\mathcal{M}Z_{0}\right]\;,\nonumber \\
c_{t} & =\left(\mu+\varepsilon\xi\right)c_{\phi}-\xi\left(F(r)-\varepsilon G(r)\right)\left[Z_{1}-\frac{1}{2}\mathcal{M}Z_{0}\right]\;,\\
c_{r} & =F(r)^{\prime}P_{2}+G(r)^{\prime}Z_{2}\;.\nonumber 
\end{align}
The spacetime metric $ds^{2}=\eta_{ab}e_{\;\mu}^{a}e_{\;\nu}^{b}dx^{\mu}dx^{\nu}$,
is then found to be given by 
\begin{equation}
ds^{2}=-f(r)^{2}\xi^{2}dt^{2}+\frac{dr^{2}}{f(r)^{2}}+r^{2}\left(d\phi+N^{\phi}(r)dt\right)^{2}\;,\label{ds2-MaxTor}
\end{equation}
where $f(r)$ is given by \eqref{f(r)-MaxTor} while $N^{\phi}(r)$
reads 
\begin{align}
N^{\phi}(r) & =\mu+\frac{1}{2}\varepsilon\xi+\frac{\mathcal{N}}{2r^{2}}\xi\;.\label{eq:Nphi-MaxTor}
\end{align}
It is reassuring that this solution corresponds to the three-dimensional
black hole solution \cite{Banados:1992wn}, which goes hand in hand
with the field equations in \eqref{Riem}. It also follows from \eqref{f(r)-MaxTor}
that for ${\cal M}>0$ together with ${\mathcal{M}+\varepsilon\mathcal{N}>0}$, this spacetime develops an inner and an event
horizon which are located at $r=r_{\pm}$, being determined according to 
\begin{equation}
r_{\pm}^{2}=\frac{1}{\varepsilon^{2}}\left(\sqrt{\mathcal{M}}\pm\sqrt{\mathcal{M}+\varepsilon\mathcal{N}}\right)^{2}\;.\label{rmasmen}
\end{equation}
In this frame the nontrivial components of spin connection are found to be given by
\begin{equation}
\omega^{0}=\frac{1}{2}{\cal M}\left(d\phi+\mu dt\right)\;,\qquad\omega^{1}=d\phi+\mu dt\;.
\end{equation}
The corresponding Maxwell CS field defined here as $\sigma_{\mu\nu}=\varepsilon\eta_{ab}e_{\;\mu}^{a}\sigma_{\;\nu}^{b}$,
is appropriately expressed as follows 
\begin{equation}
d\sigma^{2}=\varepsilon\eta_{ab}e_{\;\mu}^{a}\sigma_{\;\nu}^{b}dx^{\mu}dx^{\nu}=\left(\varrho_{\mu\nu}+\varPhi(r)\epsilon_{r\mu\nu}\right)dx^{\mu}dx^{\nu}\;,\label{dsigma2-MaxTor}
\end{equation}
with 
\begin{equation}
\varPhi(r)=\varepsilon\frac{1}{4}\xi\left[\left(\frac{\varepsilon\mathcal{F}-\mathcal{N}}{2\mathcal{M}}+\frac{1}{\varepsilon}-\frac{\nu}{\xi}\right)\left(r^{2}\varepsilon-\mathcal{N}\right)+\frac{1}{\varepsilon}\left(\varepsilon\mathcal{F}-\mathcal{N}\right)\right]\;,
\end{equation}
and where its symmetric part $\sigma_{(\mu\nu)}$ is given by 
\begin{align}
\varrho_{\mu\nu}dx^{\mu}dx^{\nu} & =\frac{\varepsilon}{2}\left[\left(1-\varepsilon\frac{\nu}{\xi}\right)\left(r^{2}\varepsilon-\mathcal{N}\right)+2\left(\mathcal{N}+\frac{\nu}{\xi}\mathcal{M}\right)\right]\xi^{2}dt^{2}+\frac{dr^{2}}{f(r)^{2}}\nonumber \\
 & +\left[r^{2}-\left(\frac{\varepsilon\mathcal{F}-\mathcal{N}}{4\mathcal{M}}\right)\left(\varepsilon r^{2}-\mathcal{N}\right)\right]\left(d\phi+\mu dt\right)^{2}\\
 & +\frac{\varepsilon}{2}\left[3r^{2}+\mathcal{F}-\left(\frac{\varepsilon\mathcal{F}-\mathcal{N}}{2\mathcal{M}}+\frac{\nu}{\xi}\right)\left(r^{2}\varepsilon-\mathcal{N}\right)\right]\xi\left(d\phi+\mu dt\right)dt\;.\nonumber 
\end{align}
The regularity of the fields around the event horizon $r=r_{+}$ can
be carried out for a fixed range of the angular coordinates of the
solid torus, i.e., $0<\tau\leq1$, and $0<\phi\leq2\pi$. The latter
can be implemented by fixing the rotating frame around $r_{+}$
in the metric \eqref{ds2-MaxTor}, which leads to choosing the Lagrange
multiplier $\mu$, according to
\begin{equation}
\mu=-\frac{1}{2}\varepsilon\xi-\frac{\mathcal{N}}{2r_{+}^{2}}\xi=-\frac{\varepsilon}{2}\left(1-\frac{r_{-}}{r_{+}}\right)\xi\;,\label{RotFr-mu-MaxTor}
\end{equation}
so that around the event horizon $r_{+}$ the metric behaves as follows
\footnote{This is carried out by defining a new radial coordinate according
to $r=\sqrt{\frac{1}{2}\left[r_{+}^{2}+r_{-}^{2}+\left(r_{+}^{2}-r_{-}^{2}\right)\text{cosh}\left(\sigma\rho\right)\right]}\,$.} 
\begin{equation}
ds^{2}\approx-\frac{\varepsilon^{4}\left(r_{+}^{2}-r_{-}^{2}\right)^{2}}{4^{2}r_{+}^{2}}\xi^{2}\rho^{2}dt^{2}+d\rho^{2}+r_{+}^{2}d\phi^{2}\;,\label{ds2-Rp-MaxTor}
\end{equation}
from which one can read that the Hawking temperature $\xi=1/T_{\text{BH}}$
is found to be 
\begin{equation}
\xi=\frac{2^{3}\pi r_{+}}{\varepsilon^{2}\left(r_{+}^{2}-r_{-}^{2}\right)}=\frac{2\pi}{|\varepsilon|}\left(\frac{1}{\sqrt{\mathcal{M}}}+\frac{1}{\sqrt{{\cal M}+\mathcal{\varepsilon\mathcal{N}}}}\right)\;.\label{Temp-MaxTor}
\end{equation}
Along the same line, expanding the Maxwell field \eqref{dsigma2-MaxTor}
around $r_{+}$ with the previous fixing \eqref{RotFr-mu-MaxTor},
it acquires the behavior 
\begin{equation}
d\sigma^{2}\approx\left[\varrho_{\mu\nu}-\frac{\varepsilon}{4}r_{+}\left(r_{+}+r_{-}\right)\left(1-\varepsilon\frac{\nu}{\xi}+\frac{\left(r_{+}-r_{-}\right)\left(\mathcal{F}+r_{+}r_{-}\right)}{r_{+}\left(r_{+}+r_{-}\right)^{2}}\right)\xi\epsilon_{r\mu\nu}\right]dx^{\mu}dx^{\nu}\;,\label{MaxF-Rp-MaxTor}
\end{equation}
where 
\begin{align}
\varrho_{\mu\nu}dx^{\mu}dx^{\nu} & \approx-\frac{\varepsilon^{4}\left(r_{+}^{2}-r_{-}^{2}\right)^{2}}{4^{2}r_{+}^{2}}\xi^{2}\rho^{2}dt^{2}+d\rho^{2}+r_{+}\left(\frac{r_{+}^{2}-\mathcal{F}}{r_{+}+r_{-}}\right)d\phi^{2}\nonumber \\
 & -\frac{\varepsilon}{4}r_{+}\left(r_{+}+r_{-}\right)\left(1-\varepsilon\frac{\nu}{\xi}+\frac{\left(r_{+}-r_{-}\right)\left(\mathcal{F}+r_{+}r_{-}\right)}{r_{+}\left(r_{+}+r_{-}\right)^{2}}\right)\xi dtd\phi\;.\label{eq:MaxF-Tor-Rc2}
\end{align}
Thus, by fixing the rotating frame in the gravitational Maxwell field
around $r_{+}$, introduces the following condition on the Lagrange multiplier $\nu$, given by 
\begin{align}
\nu & =\frac{1}{\varepsilon}\left[\frac{\left(r_{+}-r_{-}\right)\left(\mathcal{F}+r_{+}r_{-}\right)}{r_{+}\left(r_{+}+r_{-}\right)^{2}}+1\right]\xi\;.\label{RotFr-nu-MaxTor}
\end{align}
It should be emphasized that both the metric and the gravitational
Maxwell field are finite and well-behaved at the horizon $r_{+}$.
Furthermore, around the horizon both the metric \eqref{ds2-Rp-MaxTor}
and the gravitational Maxwell field \eqref{MaxF-Rp-MaxTor} by virtue
of the regularity conditions \eqref{RotFr-mu-MaxTor} and \eqref{RotFr-nu-MaxTor}
reduce to Rindler spacetimes with the same Hawking temperature determined
in \eqref{Temp-MaxTor}. It is important to note that this latter feature provides also the condition that $\varrho_{\phi\phi}\left(r_{c}\right)>0$, which would mean that both fields really share the same topology around the horizon, that is,
\begin{equation}
\varrho_{\phi\phi}\left(r_{+}\right)=r_{+}\left(\frac{r_{+}^{2}-\mathcal{F}}{r_{+}+r_{-}}\right)>0\;,
\end{equation}
which follows from \eqref{eq:MaxF-Tor-Rc2}. It is also verified that expanding the fields around the inner horizon $r_{-}$ one obtains also regularity conditions being those in  \eqref{Temp-MaxTor}, \eqref{RotFr-mu-MaxTor} and \eqref{RotFr-nu-MaxTor} by in changing $r_{\pm } \rightarrow r_{\mp}$.

In sum, in the metric formalism the regularity conditions on the fields around $r_{+}$ are found to be given by
\begin{align}
\xi & =\frac{2\pi}{|\varepsilon|}\left(\frac{1}{\sqrt{\mathcal{M}}}+\frac{1}{\sqrt{{\cal M}+\mathcal{\varepsilon\mathcal{N}}}}\right)\,, & \mu & =-\frac{2\pi}{\mathcal{M}^{1/2}}\,, & \nu & =\frac{\pi\left(\varepsilon\mathcal{F}-\mathcal{N}\right)}{\varepsilon\mathcal{M}^{3/2}}+\frac{\xi}{\varepsilon}\,.\label{RegCond-MaxTor-MetricF}
\end{align}

\bibliographystyle{fullsort.bst}
\bibliography{draft_vf}

@article{Bunster:2014mua,
    author = "Bunster, Claudio and Henneaux, Marc and Perez, Alfredo and Tempo, David and Troncoso, Ricardo",
    title = "{Generalized Black Holes in Three-dimensional Spacetime}",
    eprint = "1404.3305",
    archivePrefix = "arXiv",
    primaryClass = "hep-th",
    doi = "10.1007/JHEP05(2014)031",
    journal = "JHEP",
    volume = "05",
    pages = "031",
    year = "2014"
}

@article{Detournay:2012ug,
    author = "Detournay, Stephane",
    title = "{Inner Mechanics of 3d Black Holes}",
    eprint = "1204.6088",
    archivePrefix = "arXiv",
    primaryClass = "hep-th",
    doi = "10.1103/PhysRevLett.109.031101",
    journal = "Phys. Rev. Lett.",
    volume = "109",
    pages = "031101",
    year = "2012"
}

@article{Castro:2012av,
    author = "Castro, Alejandra and Rodriguez, Maria J.",
    title = "{Universal properties and the first law of black hole inner mechanics}",
    eprint = "1204.1284",
    archivePrefix = "arXiv",
    primaryClass = "hep-th",
    reportNumber = "NSF-KITP-12-048",
    doi = "10.1103/PhysRevD.86.024008",
    journal = "Phys. Rev. D",
    volume = "86",
    pages = "024008",
    year = "2012"
}

@article{Bagchi:2013qva,
    author = "Bagchi, Arjun and Basu, Rudranil",
    title = "{3D Flat Holography: Entropy and Logarithmic Corrections}",
    eprint = "1312.5748",
    archivePrefix = "arXiv",
    primaryClass = "hep-th",
    doi = "10.1007/JHEP03(2014)020",
    journal = "JHEP",
    volume = "03",
    pages = "020",
    year = "2014"
}

@article{Perez:2013xi,
    author = "Perez, Alfredo and Tempo, David and Troncoso, Ricardo",
    title = "{Higher spin black hole entropy in three dimensions}",
    eprint = "1301.0847",
    archivePrefix = "arXiv",
    primaryClass = "hep-th",
    reportNumber = "CECS-PHY-12-07",
    doi = "10.1007/JHEP04(2013)143",
    journal = "JHEP",
    volume = "04",
    pages = "143",
    year = "2013"
}

@article{Coussaert:1995zp,
    author = "Coussaert, Oliver and Henneaux, Marc and van Driel, Peter",
    title = "{The Asymptotic dynamics of three-dimensional Einstein gravity with a negative cosmological constant}",
    eprint = "gr-qc/9506019",
    archivePrefix = "arXiv",
    reportNumber = "ULB-TH-95-08",
    doi = "10.1088/0264-9381/12/12/012",
    journal = "Class. Quant. Grav.",
    volume = "12",
    pages = "2961--2966",
    year = "1995"
}

@article{Bagchi:2012xr,
    author = "Bagchi, Arjun and Detournay, St{\'e}phane and Fareghbal, Reza and Sim{\'o}n, Joan",
    title = "{Holography of 3D Flat Cosmological Horizons}",
    eprint = "1208.4372",
    archivePrefix = "arXiv",
    primaryClass = "hep-th",
    reportNumber = "EMPG-12-18",
    doi = "10.1103/PhysRevLett.110.141302",
    journal = "Phys. Rev. Lett.",
    volume = "110",
    number = "14",
    pages = "141302",
    year = "2013"
}

@article{Henneaux:2013dra,
    author = "Henneaux, Marc and Perez, Alfredo and Tempo, David and Troncoso, Ricardo",
    title = "{Chemical potentials in three-dimensional higher spin anti-de Sitter gravity}",
    eprint = "1309.4362",
    archivePrefix = "arXiv",
    primaryClass = "hep-th",
    doi = "10.1007/JHEP12(2013)048",
    journal = "JHEP",
    volume = "12",
    pages = "048",
    year = "2013"
}

@article{Blagojevic:2013aaa,
    author = "Blagojevi{\'c}, M. and Cvetkovi{\'c}, B. and Vasili{\'c}, M.",
    title = "{{\textquotedblleft}Exotic{\textquotedblright} black holes with torsion}",
    eprint = "1310.1412",
    archivePrefix = "arXiv",
    primaryClass = "gr-qc",
    doi = "10.1103/PhysRevD.88.101501",
    journal = "Phys. Rev. D",
    volume = "88",
    number = "10",
    pages = "101501",
    year = "2013"
}

@article{Matulich:2023xpw,
    author = "Matulich, Javier and Rodr{\'\i}guez, Evelyn",
    title = "{Enlarged super-bms$_3$ algebra and its flat limit}",
    eprint = "2310.16614",
    archivePrefix = "arXiv",
    primaryClass = "hep-th",
    doi = "10.1103/PhysRevD.110.064064",
    journal = "Phys. Rev. D",
    volume = "110",
    number = "6",
    pages = "064064",
    year = "2024"
}

@article{Witten:1988hc,
      author         = "Witten, Edward",
      title          = "{(2+1)-Dimensional Gravity as an Exactly Soluble System}",
      journal        = "Nucl. Phys.",
      volume         = "B311",
      year           = "1988",
      pages          = "46",
      doi            = "10.1016/0550-3213(88)90143-5",
      reportNumber   = "IASSNS-HEP-88-32",
      SLACcitation   = "%%CITATION = NUPHA,B311,46;%%"
}

@article{Ashtekar:1996cd,
      author         = "Ashtekar, Abhay and Bičák, Jiří and Schmidt, Bernd G",
      title          = "{Asymptotic structure of symmetry reduced general
                        relativity}",
      journal        = "Phys. Rev.",
      volume         = "D55",
      year           = "1997",
      pages          = "669-686",
      doi            = "10.1103/PhysRevD.55.669",
      eprint         = "gr-qc/9608042",
      archivePrefix  = "arXiv",
      primaryClass   = "gr-qc",
      reportNumber   = "CGPG-96-5-1",
      SLACcitation   = "%%CITATION = GR-QC/9608042;%%"
}

@article{Barnich:2006av,
      author         = "Barnich, Glenn and Compere, Geoffrey",
      title          = "{Classical central extension for asymptotic symmetries at
                        null infinity in three spacetime dimensions}",
      journal        = "Class. Quant. Grav.",
      volume         = "24",
      year           = "2007",
      pages          = "F15-F23",
      doi            = "10.1088/0264-9381/24/5/F01, 10.1088/0264-9381/24/11/C01",
      eprint         = "gr-qc/0610130",
      archivePrefix  = "arXiv",
      primaryClass   = "gr-qc",
      reportNumber   = "ULB-TH-06-08",
      SLACcitation   = "%%CITATION = GR-QC/0610130;%%"
}

@article{Banados:1994tn,
      author         = "Banados, Maximo",
      title          = "{Global charges in Chern-Simons field theory and the
                        (2+1) black hole}",
      journal        = "Phys. Rev.",
      volume         = "D52",
      year           = "1996",
      pages          = "5816-5825",
      doi            = "10.1103/PhysRevD.52.5816",
      eprint         = "hep-th/9405171",
      archivePrefix  = "arXiv",
      primaryClass   = "hep-th",
      reportNumber   = "IMPERIAL-TP-93-94-40",
      SLACcitation   = "%%CITATION = HEP-TH/9405171;%%"
}

@article{Solodukhin:2005ah,
    author = "Solodukhin, Sergey N.",
    title = "{Holography with gravitational Chern-Simons}",
    eprint = "hep-th/0509148",
    archivePrefix = "arXiv",
    reportNumber = "IUB-TH-057",
    doi = "10.1103/PhysRevD.74.024015",
    journal = "Phys. Rev. D",
    volume = "74",
    pages = "024015",
    year = "2006"
}

@article{Cangemi:1992ri,
    author = "Cangemi, D.",
    title = "{One formulation for both lineal gravities through a dimensional reduction}",
    eprint = "gr-qc/9207004",
    archivePrefix = "arXiv",
    reportNumber = "MIT-CTP-2124, CTP{\#}2124",
    doi = "10.1016/0370-2693(92)91259-C",
    journal = "Phys. Lett. B",
    volume = "297",
    pages = "261--265",
    year = "1992"
}

@article{Fukuyama:1985gg,
    author = "Fukuyama, Takeshi and Kamimura, Kiyoshi",
    title = "{Gauge Theory of Two-dimensional Gravity}",
    reportNumber = "Print-85-0318 (TOHO)",
    doi = "10.1016/0370-2693(85)91322-X",
    journal = "Phys. Lett. B",
    volume = "160",
    pages = "259--262",
    year = "1985"
}

@article{Isler:1989hq,
    author = "Isler, K. and Trugenberger, C. A.",
    title = "{A Gauge Theory of Two-dimensional Quantum Gravity}",
    reportNumber = "MIT-CTP-1739",
    doi = "10.1103/PhysRevLett.63.834",
    journal = "Phys. Rev. Lett.",
    volume = "63",
    pages = "834",
    year = "1989"
}

@article{Chamseddine:1989yz,
    author = "Chamseddine, Ali H. and Wyler, D.",
    title = "{Gauge Theory of Topological Gravity in (1+1)-Dimensions}",
    reportNumber = "Print-89-0614 (ZURICH)",
    doi = "10.1016/0370-2693(89)90528-5",
    journal = "Phys. Lett. B",
    volume = "228",
    pages = "75--78",
    year = "1989"
}

@article{Cangemi:1992bj,
    author = "Cangemi, D. and Jackiw, R.",
    title = "{Gauge invariant formulations of lineal gravity}",
    eprint = "hep-th/9203056",
    archivePrefix = "arXiv",
    reportNumber = "MIT-CTP-2085",
    doi = "10.1103/PhysRevLett.69.233",
    journal = "Phys. Rev. Lett.",
    volume = "69",
    pages = "233--236",
    year = "1992"
}

@article{Cangemi:1993sd,
    author = "Cangemi, D. and Jackiw, R.",
    title = "{Poincare gauge theory for gravitational forces in (1+1)-dimensions}",
    eprint = "hep-th/9302026",
    archivePrefix = "arXiv",
    reportNumber = "MIT-CTP-2165",
    doi = "10.1006/aphy.1993.1058",
    journal = "Annals Phys.",
    volume = "225",
    pages = "229--263",
    year = "1993"
}

@article{Bonora:2011gz,
    author = "Bonora, L. and Cvitan, M. and Dominis Prester, P. and Pallua, S. and Smolic, I.",
    title = "{Gravitational Chern-Simons Lagrangians and black hole entropy}",
    eprint = "1104.2523",
    archivePrefix = "arXiv",
    primaryClass = "hep-th",
    reportNumber = "SISSA-89-2010-EP, YITP-10-113, ZTF-11-03",
    doi = "10.1007/JHEP07(2011)085",
    journal = "JHEP",
    volume = "07",
    pages = "085",
    year = "2011"
}

@article{Banados:1992wn,
      author         = "Banados, Maximo and Teitelboim, Claudio and Zanelli,
                        Jorge",
      title          = "{The Black hole in three-dimensional space-time}",
      journal        = "Phys. Rev. Lett.",
      volume         = "69",
      year           = "1992",
      pages          = "1849-1851",
      doi            = "10.1103/PhysRevLett.69.1849",
      eprint         = "hep-th/9204099",
      archivePrefix  = "arXiv",
      primaryClass   = "hep-th",
      reportNumber   = "PRINT-92-0151 (CHILE), IASSNS-HEP-92-29",
      SLACcitation   = "%%CITATION = HEP-TH/9204099;%%"
}

@article{Hoseinzadeh:2014bla,
    author = "Hoseinzadeh, S. and Rezaei-Aghdam, A.",
    title = "{(2$+$1)-dimensional gravity from Maxwell and semisimple extension of the Poincar\'e gauge symmetric models}",
    eprint = "1402.0320",
    archivePrefix = "arXiv",
    primaryClass = "hep-th",
    doi = "10.1103/PhysRevD.90.084008",
    journal = "Phys. Rev. D",
    volume = "90",
    number = "8",
    pages = "084008",
    year = "2014"
}

@article{Barnich:2012xq,
    author = "Barnich, Glenn",
    title = "{Entropy of three-dimensional asymptotically flat cosmological solutions}",
    eprint = "1208.4371",
    archivePrefix = "arXiv",
    primaryClass = "hep-th",
    reportNumber = "ULB-TH-12-14",
    doi = "10.1007/JHEP10(2012)095",
    journal = "JHEP",
    volume = "10",
    pages = "095",
    year = "2012"
}

@article{Caroca:2021bjo,
    author = "Caroca, Ricardo and Concha, Patrick and Matulich, Javier and Rodr{\'\i}guez, Evelyn and Tempo, David",
    title = "{Hypersymmetric extensions of Maxwell-Chern-Simons gravity in 2+1 dimensions}",
    eprint = "2105.12243",
    archivePrefix = "arXiv",
    primaryClass = "hep-th",
    doi = "10.1103/PhysRevD.104.064011",
    journal = "Phys. Rev. D",
    volume = "104",
    number = "6",
    pages = "064011",
    year = "2021"
}

@article{Duval:2008tr,
    author = "Duval, C. and Horvath, Z. and Horvathy, P. A.",
    title = "{Chern-Simons gravity, based on a non-semisimple group}",
    eprint = "0807.0977",
    archivePrefix = "arXiv",
    primaryClass = "hep-th",
    month = "7",
    year = "2008"
}

@article{Hofenstock:2025xoc,
    author = {H{\"o}fenstock, Felix and Salgado-Rebolledo, Patricio},
    title = "{Boundary dynamics of Maxwell-invariant three-dimensional Chern-Simons gravity}",
    eprint = "2506.07651",
    archivePrefix = "arXiv",
    primaryClass = "hep-th",
    month = "6",
    year = "2025"
}

@article{Salgado-Rebolledo:2019kft,
    author = "Salgado-Rebolledo, Patricio",
    title = "{The Maxwell group in 2+1 dimensions and its infinite-dimensional enhancements}",
    eprint = "1905.09421",
    archivePrefix = "arXiv",
    primaryClass = "hep-th",
    doi = "10.1007/JHEP10(2019)039",
    journal = "JHEP",
    volume = "10",
    pages = "039",
    year = "2019"
}

@article{Aviles:2023igk,
    author = "Avil{\'e}s, Luis and Hidalgo, Diego and Valdivia, Omar",
    title = "{Thermodynamics of the three-dimensional black hole with torsion}",
    eprint = "2308.09114",
    archivePrefix = "arXiv",
    primaryClass = "gr-qc",
    doi = "10.1007/JHEP09(2023)185",
    journal = "JHEP",
    volume = "09",
    pages = "185",
    year = "2023"
}

@article{schrader1972maxwell,
  title={The Maxwell group and the quantum theory of particles in classical homogeneous electromagnetic fields},
  author={Schrader, Robert},
  journal={Fortschritte der Physik},
  volume={20},
  number={12},
  pages={701--734},
  year={1972},
  publisher={Wiley Online Library}
}

@article{Grumiller:2016kcp,
      author         = "Grumiller, Daniel and Perez, Alfredo and Prohazka, Stefan
                        and Tempo, David and Troncoso, Ricardo",
      title          = "{Higher Spin Black Holes with Soft Hair}",
      journal        = "JHEP",
      volume         = "10",
      year           = "2016",
      pages          = "119",
      doi            = "10.1007/JHEP10(2016)119",
      eprint         = "1607.05360",
      archivePrefix  = "arXiv",
      primaryClass   = "hep-th",
      reportNumber   = "CECS-PHY-16-03, TUW-16-13",
      SLACcitation   = "%%CITATION = ARXIV:1607.05360;%%"
}

@article{Caroca:2017onr,
      author         = "Caroca, Ricardo and Concha, Patrick and Rodríguez,
                        Evelyn and Salgado-Rebolledo, Patricio",
      title          = "{Generalizing the $\mathfrak {bms}_{3}$ and 2D-conformal
                        algebras by expanding the Virasoro algebra}",
      journal        = "Eur. Phys. J.",
      volume         = "C78",
      year           = "2018",
      number         = "3",
      pages          = "262",
      doi            = "10.1140/epjc/s10052-018-5739-7",
      eprint         = "1707.07209",
      archivePrefix  = "arXiv",
      primaryClass   = "hep-th",
      reportNumber   = "UAI-PHY-17-05",
      SLACcitation   = "%%CITATION = ARXIV:1707.07209;%%"
}

@article{Concha:2018zeb,
      author         = "Concha, Patrick and Merino, Nelson and Miskovic, Olivera
                        and Rodríguez, Evelyn and Salgado-Rebolledo, Patricio and
                        Valdivia, Omar",
      title          = "{Asymptotic symmetries of three-dimensional Chern-Simons
                        gravity for the Maxwell algebra}",
      journal        = "JHEP",
      volume         = "10",
      year           = "2018",
      pages          = "079",
      doi            = "10.1007/JHEP10(2018)079",
      eprint         = "1805.08834",
      archivePrefix  = "arXiv",
      primaryClass   = "hep-th",
      reportNumber   = "UAI-PHY-18/05, UAI-PHY-18-05",
      SLACcitation   = "%%CITATION = ARXIV:1805.08834;%%"
}

@article{deAzcarraga:2012qj,
      author         = "de Azcarraga, Jose A. and Kamimura, Kiyoshi and
                        Lukierski, Jerzy",
      title          = "{Maxwell symmetries and some applications}",
      booktitle      = "{Proceedings, 3rd Galileo - Xu Guangqi Meeting on The
                        Sun, the Stars. the Universe and General Relativity (GX3):
                        Beijing, China, October 11-15, 2011}",
      journal        = "Int. J. Mod. Phys. Conf. Ser.",
      volume         = "23",
      year           = "2013",
      pages          = "01160",
      doi            = "10.1142/S2010194513011604",
      eprint         = "1201.2850",
      archivePrefix  = "arXiv",
      primaryClass   = "hep-th",
      reportNumber   = "IFIC-12-01, FTUV-12-0113, IFIC-12-01---FTUV-12-0113",
      SLACcitation   = "%%CITATION = ARXIV:1201.2850;%%"
}

@article{Bacry:1970ye,
      author         = "Bacry, H. and Combe, P. and Richard, J. L.",
      title          = "{Group-theoretical analysis of elementary particles in an
                        external electromagnetic field. 1. the relativistic
                        particle in a constant and uniform field}",
      journal        = "Nuovo Cim.",
      volume         = "A67",
      year           = "1970",
      pages          = "267-299",
      doi            = "10.1007/BF02725178",
      SLACcitation   = "%%CITATION = NUCIA,A67,267;%%"
}

@article{deAzcarraga:2010sw,
      author         = "de Azcarraga, Jose A. and Kamimura, Kiyoshi and
                        Lukierski, Jerzy",
      title          = "{Generalized cosmological term from Maxwell symmetries}",
      journal        = "Phys. Rev.",
      volume         = "D83",
      year           = "2011",
      pages          = "124036",
      doi            = "10.1103/PhysRevD.83.124036",
      eprint         = "1012.4402",
      archivePrefix  = "arXiv",
      primaryClass   = "hep-th",
      SLACcitation   = "%%CITATION = ARXIV:1012.4402;%%"
}

@article{Izaurieta:2009hz,
      author         = "Izaurieta, Fernando and Rodriguez, Eduardo and Minning,
                        Paul and Salgado, Patricio and Perez, Alfredo",
      title          = "{Standard General Relativity from Chern-Simons Gravity}",
      journal        = "Phys. Lett.",
      volume         = "B678",
      year           = "2009",
      pages          = "213-217",
      doi            = "10.1016/j.physletb.2009.06.017",
      eprint         = "0905.2187",
      archivePrefix  = "arXiv",
      primaryClass   = "hep-th",
      SLACcitation   = "%%CITATION = ARXIV:0905.2187;%%"
}

@article{Concha:2013uhq,
      author         = "Concha, P. K. and Peñafiel, D. M. and Rodríguez, E. K.
                        and Salgado, P.",
      title          = "{Even-dimensional General Relativity from Born-Infeld
                        gravity}",
      journal        = "Phys. Lett.",
      volume         = "B725",
      year           = "2013",
      pages          = "419-424",
      doi            = "10.1016/j.physletb.2013.07.019",
      eprint         = "1309.0062",
      archivePrefix  = "arXiv",
      primaryClass   = "hep-th",
      SLACcitation   = "%%CITATION = ARXIV:1309.0062;%%"
}

@article{Salgado:2014jka,
      author         = "Salgado, Patricio and Szabo, Richard J. and Valdivia,
                        Omar",
      title          = "{Topological gravity and transgression holography}",
      journal        = "Phys. Rev.",
      volume         = "D89",
      year           = "2014",
      number         = "8",
      pages          = "084077",
      doi            = "10.1103/PhysRevD.89.084077",
      eprint         = "1401.3653",
      archivePrefix  = "arXiv",
      primaryClass   = "hep-th",
      reportNumber   = "EMPG-14-01",
      SLACcitation   = "%%CITATION = ARXIV:1401.3653;%%"
}

@article{Concha:2018jjj,
      author         = "Concha, Patrick and Merino, Nelson and Rodríguez, Evelyn
                        and Salgado-Rebolledo, Patricio and Valdivia, Omar",
      title          = "{Semi-simple enlargement of the $\mathfrak{bms}_3$
                        algebra from a
                        $\mathfrak{so}(2,2)\oplus\mathfrak{so}(2,1)$ Chern-Simons
                        theory}",
      journal        = "JHEP",
      volume         = "02",
      year           = "2019",
      pages          = "002",
      doi            = "10.1007/JHEP02(2019)002",
      eprint         = "1810.12256",
      archivePrefix  = "arXiv",
      primaryClass   = "hep-th",
      SLACcitation   = "%%CITATION = ARXIV:1810.12256;%%"
}

@article{Soroka:2006aj,
      author         = "Soroka, Dmitrij V. and Soroka, Vyacheslav A.",
      title          = "{Semi-simple extension of the (super)Poincare algebra}",
      journal        = "Adv. High Energy Phys.",
      volume         = "2009",
      year           = "2009",
      pages          = "234147",
      doi            = "10.1155/2009/234147",
      eprint         = "hep-th/0605251",
      archivePrefix  = "arXiv",
      primaryClass   = "hep-th",
      SLACcitation   = "%%CITATION = HEP-TH/0605251;%%"
}

@article{Durka:2011nf,
      author         = "Durka, R. and Kowalski-Glikman, J. and Szczachor, M.",
      title          = "{Gauged AdS-Maxwell algebra and gravity}",
      journal        = "Mod. Phys. Lett.",
      volume         = "A26",
      year           = "2011",
      pages          = "2689-2696",
      doi            = "10.1142/S0217732311037078",
      eprint         = "1107.4728",
      archivePrefix  = "arXiv",
      primaryClass   = "hep-th",
      SLACcitation   = "%%CITATION = ARXIV:1107.4728;%%"
}

@article{Aviles:2018jzw,
      author         = "Avilés, Luis and Frodden, Ernesto and Gomis, Joaquim and
                        Hidalgo, Diego and Zanelli, Jorge",
      title          = "{Non-Relativistic Maxwell Chern-Simons Gravity}",
      journal        = "JHEP",
      volume         = "05",
      year           = "2018",
      pages          = "047",
      doi            = "10.1007/JHEP05(2018)047",
      eprint         = "1802.08453",
      archivePrefix  = "arXiv",
      primaryClass   = "hep-th",
      SLACcitation   = "%%CITATION = ARXIV:1802.08453;%%"
}

@article{Salgado:2014qqa,
      author         = "Salgado, P. and Salgado, S.",
      title          = "{$\mathfrak{so}(D-1,1)\otimes \mathfrak{so}(D-1,2)$
                        algebras and gravity}",
      journal        = "Phys. Lett.",
      volume         = "B728",
      year           = "2014",
      pages          = "5-10",
      doi            = "10.1016/j.physletb.2013.11.009",
      SLACcitation   = "%%CITATION = PHLTA,B728,5;%%"
}

@article{Matulich:2014hea,
      author         = "Matulich, Javier and Perez, Alfredo and Tempo, David and
                        Troncoso, Ricardo",
      title          = "{Higher spin extension of cosmological spacetimes in 3D:
                        asymptotically flat behaviour with chemical potentials and
                        thermodynamics}",
      journal        = "JHEP",
      volume         = "05",
      year           = "2015",
      pages          = "025",
      doi            = "10.1007/JHEP05(2015)025",
      eprint         = "1412.1464",
      archivePrefix  = "arXiv",
      primaryClass   = "hep-th",
      reportNumber   = "CECS-PHY-14-03",
      SLACcitation   = "%%CITATION = ARXIV:1412.1464;%%"
}

@article{Concha:2014vka,
      author         = "Concha, P. K. and Penafiel, D. M. and Rodriguez, E. K.
                        and Salgado, P.",
      title          = "{Chern-Simons and Born-Infeld gravity theories and
                        Maxwell algebras type}",
      journal        = "Eur. Phys. J.",
      volume         = "C74",
      year           = "2014",
      pages          = "2741",
      doi            = "10.1140/epjc/s10052-014-2741-6",
      eprint         = "1402.0023",
      archivePrefix  = "arXiv",
      primaryClass   = "hep-th",
      SLACcitation   = "%%CITATION = ARXIV:1402.0023;%%"
}

@article{Caroca:2017izc,
      author         = "Caroca, Ricardo and Concha, Patrick and Fierro, Octavio
                        and Rodríguez, Evelyn and Salgado-Rebolledo, Patricio",
      title          = "{Generalized Chern–Simons higher-spin gravity theories
                        in three dimensions}",
      journal        = "Nucl. Phys.",
      volume         = "B934",
      year           = "2018",
      pages          = "240-264",
      doi            = "10.1016/j.nuclphysb.2018.07.005",
      eprint         = "1712.09975",
      archivePrefix  = "arXiv",
      primaryClass   = "hep-th",
      reportNumber   = "UAI-PHY-18/03, UAI-PHY-18-03",
      SLACcitation   = "%%CITATION = ARXIV:1712.09975;%%"
}

@article{Concha:2014tca,
      author         = "Concha, P. K. and Rodríguez, E. K.",
      title          = "{N = 1 Supergravity and Maxwell superalgebras}",
      journal        = "JHEP",
      volume         = "09",
      year           = "2014",
      pages          = "090",
      doi            = "10.1007/JHEP09(2014)090",
      eprint         = "1407.4635",
      archivePrefix  = "arXiv",
      primaryClass   = "hep-th",
      SLACcitation   = "%%CITATION = ARXIV:1407.4635;%%"
}

@article{Ravera:2018vra,
      author         = "Ravera, Lucrezia",
      title          = "{Hidden role of Maxwell superalgebras in the free
                        differential algebras of D = 4 and D = 11
                        supergravity}",
      journal        = "Eur. Phys. J.",
      volume         = "C78",
      year           = "2018",
      number         = "3",
      pages          = "211",
      doi            = "10.1140/epjc/s10052-018-5673-8",
      eprint         = "1801.08860",
      archivePrefix  = "arXiv",
      primaryClass   = "hep-th",
      SLACcitation   = "%%CITATION = ARXIV:1801.08860;%%"
}

@article{Concha:2018jxx,
      author         = "Concha, Patrick and Peñafiel, Diego M. and Rodríguez,
                        Evelyn",
      title          = "{On the Maxwell supergravity and flat limit in 2 + 1
                        dimensions}",
      journal        = "Phys. Lett.",
      volume         = "B785",
      year           = "2018",
      pages          = "247-253",
      doi            = "10.1016/j.physletb.2018.08.050",
      eprint         = "1807.00194",
      archivePrefix  = "arXiv",
      primaryClass   = "hep-th",
      SLACcitation   = "%%CITATION = ARXIV:1807.00194;%%"
}

@article{Concha:2019eip,
    author = "Concha, Patrick and Safari, H.R.",
    archivePrefix = "arXiv",
    doi = "10.1007/JHEP04(2020)073",
    eprint = "1909.12827",
    journal = "JHEP",
    pages = "073",
    primaryClass = "hep-th",
    reportNumber = "IPM/P-2019/037",
    title = "{On Stabilization of Maxwell-BMS Algebra}",
    volume = "04",
    year = "2020"
}

@article{Banados:1992gq,
    author = "Banados, Maximo and Henneaux, Marc and Teitelboim, Claudio and Zanelli, Jorge",
    title = "{Geometry of the (2+1) black hole}",
    eprint = "gr-qc/9302012",
    archivePrefix = "arXiv",
    reportNumber = "IASSNS-HEP-92-81",
    doi = "10.1103/PhysRevD.48.1506",
    journal = "Phys. Rev. D",
    volume = "48",
    pages = "1506--1525",
    year = "1993",
    note = "[Erratum: Phys.Rev.D 88, 069902 (2013)]"
}

@article{Banados:1993ur,
    author = "Banados, Maximo and Teitelboim, Claudio and Zanelli, Jorge",
    title = "{Dimensionally continued black holes}",
    eprint = "gr-qc/9307033",
    archivePrefix = "arXiv",
    reportNumber = "IASSNS-AST-93-45",
    doi = "10.1103/PhysRevD.49.975",
    journal = "Phys. Rev. D",
    volume = "49",
    pages = "975--986",
    year = "1994"
}

@article{Crisostomo:2000bb,
    author = "Crisostomo, Juan and Troncoso, Ricardo and Zanelli, Jorge",
    title = "{Black hole scan}",
    eprint = "hep-th/0003271",
    archivePrefix = "arXiv",
    reportNumber = "CECS-PHY-00-01, ULB-TH-00-01",
    doi = "10.1103/PhysRevD.62.084013",
    journal = "Phys. Rev. D",
    volume = "62",
    pages = "084013",
    year = "2000"
}

@article{REGGE1974286,
title = "Role of surface integrals in the Hamiltonian formulation of general relativity",
journal = "Annals of Physics",
volume = "88",
number = "1",
pages = "286 - 318",
year = "1974",
issn = "0003-4916",
doi = "https://doi.org/10.1016/0003-4916(74)90404-7",
url = "http://www.sciencedirect.com/science/article/pii/0003491674904047",
author = "Tullio Regge and Claudio Teitelboim"
}

@inproceedings{Zanelli:2005sa,
    author = "Zanelli, Jorge",
    archivePrefix = "arXiv",
    booktitle = "{Geometric and topological methods for quantum field theory. Proceedings, Summer School, Villa de Leyva, Colombia, July 9-27, 2001}",
    eprint = "hep-th/0502193",
    month = "2",
    title = "{Lecture notes on Chern-Simons (super-)gravities. Second edition (February 2008)}",
    year = "2005"
}

@article{Barnich:2010eb,
    author = "Barnich, Glenn and Troessaert, Cedric",
    archivePrefix = "arXiv",
    doi = "10.1007/JHEP05(2010)062",
    eprint = "1001.1541",
    journal = "JHEP",
    pages = "062",
    primaryClass = "hep-th",
    reportNumber = "ULB-TH-09-28",
    title = "{Aspects of the BMS/CFT correspondence}",
    volume = "05",
    year = "2010"
}

@article{Gomis:2017cmt,
    author = "Gomis, Joaquim and Kleinschmidt, Axel",
    archivePrefix = "arXiv",
    doi = "10.1007/JHEP07(2017)085",
    eprint = "1705.05854",
    journal = "JHEP",
    pages = "085",
    primaryClass = "hep-th",
    reportNumber = "ICCUB-17--002",
    title = "{On free Lie algebras and particles in electro-magnetic fields}",
    volume = "07",
    year = "2017"
}

@article{Edelstein:2006se,
    author = "Edelstein, Jose D. and Hassaine, Mokhtar and Troncoso, Ricardo and Zanelli, Jorge",
    archivePrefix = "arXiv",
    doi = "10.1016/j.physletb.2006.07.058",
    eprint = "hep-th/0605174",
    journal = "Phys. Lett. B",
    pages = "278--284",
    reportNumber = "CECS-PHY-06-09",
    title = "{Lie-algebra expansions, Chern-Simons theories and the Einstein-Hilbert Lagrangian}",
    volume = "640",
    year = "2006"
}

@article{Chernyavsky:2020fqs,
    author = "Chernyavsky, Dmitry and Deger, Nihat Sadik and Sorokin, Dmitri",
    archivePrefix = "arXiv",
    eprint = "2002.07592",
    month = "2",
    primaryClass = "hep-th",
    title = "{Spontaneously Broken 3$d$ Hietarinta-Maxwell Chern-Simons Theory and Minimal Massive Gravity}",
    year = "2020"
}

@article{Concha:2019mxx,
    author = "Concha, Patrick and Ravera, Lucrezia and Rodríguez, Evelyn",
    archivePrefix = "arXiv",
    doi = "10.1007/JHEP04(2020)051",
    eprint = "1912.09477",
    journal = "JHEP",
    pages = "051",
    primaryClass = "hep-th",
    title = "{Three-dimensional Maxwellian extended Bargmann supergravity}",
    volume = "04",
    year = "2020"
}

@article{Concha:2018ywv,
    author = "Concha, Patrick and Ravera, Lucrezia and Rodr{\'\i}guez, Evelyn",
    title = "{On the supersymmetry invariance of flat supergravity with boundary}",
    eprint = "1809.07871",
    archivePrefix = "arXiv",
    primaryClass = "hep-th",
    doi = "10.1007/JHEP01(2019)192",
    journal = "JHEP",
    volume = "01",
    pages = "192",
    year = "2019"
}

@article{Blagojevic:2003uc,
    author = "Blagojevic, M. and Vasilic, M.",
    archivePrefix = "arXiv",
    doi = "10.1103/PhysRevD.67.084032",
    eprint = "gr-qc/0301051",
    journal = "Phys. Rev. D",
    pages = "084032",
    title = "{Asymptotic symmetries in 3-d gravity with torsion}",
    volume = "67",
    year = "2003"
}

@article{Blagojevic:2003vn,
    author = "Blagojevic, M. and Vasilic, M.",
    archivePrefix = "arXiv",
    doi = "10.1103/PhysRevD.68.104023",
    eprint = "gr-qc/0307078",
    journal = "Phys. Rev. D",
    pages = "104023",
    title = "{3-D gravity with torsion as a Chern-Simons gauge theory}",
    volume = "68",
    year = "2003"
}

@article{Blagojevic:2006jk,
    author = "Blagojevic, M. and Cvetkovic, B.",
    archivePrefix = "arXiv",
    doi = "10.1088/0264-9381/23/14/013",
    eprint = "gr-qc/0601006",
    journal = "Class. Quant. Grav.",
    pages = "4781",
    title = "{Black hole entropy in 3-D gravity with torsion}",
    volume = "23",
    year = "2006"
}

@article{Blagojevic:2006hh,
    author = "Blagojevic, M. and Cvetkovic, B.",
    archivePrefix = "arXiv",
    doi = "10.1088/1126-6708/2006/10/005",
    eprint = "gr-qc/0606086",
    journal = "JHEP",
    pages = "005",
    title = "{Black hole entropy from the boundary conformal structure in 3D gravity with torsion}",
    volume = "10",
    year = "2006"
}

@article{Giacomini:2006dr,
    author = "Giacomini, Alex and Troncoso, Ricardo and Willison, Steven",
    archivePrefix = "arXiv",
    doi = "10.1088/0264-9381/24/11/005",
    eprint = "hep-th/0610077",
    journal = "Class. Quant. Grav.",
    pages = "2845--2860",
    title = "{Three-dimensional supergravity reloaded}",
    volume = "24",
    year = "2007"
}

@article{Garcia:2003nm,
    author = "Garcia, Alberto A. and Hehl, Friedrich W. and Heinicke, Christian and Macias, Alfredo",
    title = "{Exact vacuum solution of a (1+2)-dimensional Poincare gauge theory: BTZ solution with torsion}",
    eprint = "gr-qc/0302097",
    archivePrefix = "arXiv",
    doi = "10.1103/PhysRevD.67.124016",
    journal = "Phys. Rev. D",
    volume = "67",
    pages = "124016",
    year = "2003"
}

@article{Mielke:2003xx,
    author = "Mielke, Eckehard W. and Rincon Maggiolo, Ali A.",
    title = "{Rotating black hole solution in a generalized topological 3-D gravity with torsion}",
    doi = "10.1103/PhysRevD.68.104026",
    journal = "Phys. Rev. D",
    volume = "68",
    pages = "104026",
    year = "2003"
}

@article{Blagojevic:2006nf,
    author = "Blagojevic, M. and Cvetkovic, B.",
    title = "{Covariant description of the black hole entropy in 3D gravity}",
    eprint = "gr-qc/0607026",
    archivePrefix = "arXiv",
    doi = "10.1088/0264-9381/24/1/007",
    journal = "Class. Quant. Grav.",
    volume = "24",
    pages = "129--140",
    year = "2007"
}

@article{Adami:2020xkm,
    author = "Adami, H. and Concha, P. and Rodriguez, E. and Safari, H.R.",
    title = "{Asymptotic Symmetries of Maxwell Chern-Simons Gravity with Torsion}",
    eprint = "2005.07690",
    archivePrefix = "arXiv",
    primaryClass = "hep-th",
    doi = "10.1140/epjc/s10052-020-08537-z",
    journal = "Eur. Phys. J. C",
    volume = "80",
    number = "10",
    pages = "967",
    year = "2020"
}

@article{Concha:2024rac,
    author = "Concha, Patrick and Matulich, Javier and Pino, Daniel and Rodr{\'\i}guez, Evelyn",
    title = "{Asymptotic structure of three-dimensional Maxwell Chern-Simons gravity coupled to spin-3 fields}",
    eprint = "2412.04992",
    archivePrefix = "arXiv",
    primaryClass = "hep-th",
    doi = "10.1007/JHEP02(2025)148",
    journal = "JHEP",
    volume = "02",
    pages = "148",
    year = "2025"
}

@article{Caroca:2021njq,
    author = "Caroca, Ricardo and Concha, Patrick and Pe{\~n}afiel, Diego and Rodr{\'\i}guez, Evelyn",
    title = "{Three-dimensional teleparallel Chern-Simons supergravity theory}",
    eprint = "2103.06717",
    archivePrefix = "arXiv",
    primaryClass = "hep-th",
    doi = "10.1140/epjc/s10052-021-09554-2",
    journal = "Eur. Phys. J. C",
    volume = "81",
    number = "8",
    pages = "762",
    year = "2021"
}

@article{Aviles:2025ygw,
    author = "Avil{\'e}s, Luis and Fuentealba, Oscar and Hidalgo, Diego and Rodr{\'\i}guez, Pablo",
    title = "{AdS$_{3}$ Carroll gravity: asymptotic symmetries and C-thermal configurations}",
    eprint = "2503.18818",
    archivePrefix = "arXiv",
    primaryClass = "hep-th",
    doi = "10.1007/JHEP05(2025)174",
    journal = "JHEP",
    volume = "05",
    pages = "174",
    year = "2025"
}

@article{Christensen:2013lma,
    author = "Christensen, Morten H. and Hartong, Jelle and Obers, Niels A. and Rollier, Blaise",
    title = "{Torsional Newton-Cartan Geometry and Lifshitz Holography}",
    eprint = "1311.4794",
    archivePrefix = "arXiv",
    primaryClass = "hep-th",
    doi = "10.1103/PhysRevD.89.061901",
    journal = "Phys. Rev. D",
    volume = "89",
    pages = "061901",
    year = "2014"
}

@article{Geracie:2015dea,
    author = "Geracie, Michael and Prabhu, Kartik and Roberts, Matthew M.",
    title = "{Curved non-relativistic spacetimes, Newtonian gravitation and massive matter}",
    eprint = "1503.02682",
    archivePrefix = "arXiv",
    primaryClass = "hep-th",
    reportNumber = "EFI-15-14",
    doi = "10.1063/1.4932967",
    journal = "J. Math. Phys.",
    volume = "56",
    number = "10",
    pages = "103505",
    year = "2015"
}

@article{Concha:2020ebl,
    author = "Concha, Patrick and Ravera, Lucrezia and Rodr{\'\i}guez, Evelyn and Rubio, Gustavo",
    title = "{Three-dimensional Maxwellian Extended Newtonian gravity and flat limit}",
    eprint = "2006.13128",
    archivePrefix = "arXiv",
    primaryClass = "hep-th",
    doi = "10.1007/JHEP10(2020)181",
    journal = "JHEP",
    volume = "10",
    pages = "181",
    year = "2020"
}

@article{Concha:2020eam,
    author = "Concha, Patrick and Ipinza, Marcelo and Ravera, Lucrezia and Rodr{\'\i}guez, Evelyn",
    title = "{Non-relativistic three-dimensional supergravity theories and semigroup expansion method}",
    eprint = "2010.01216",
    archivePrefix = "arXiv",
    primaryClass = "hep-th",
    doi = "10.1007/JHEP02(2021)094",
    journal = "JHEP",
    volume = "02",
    pages = "094",
    year = "2021"
}

@article{Concha:2021jnn,
    author = "Concha, Patrick and Pe{\~n}afiel, Diego and Ravera, Lucrezia and Rodr{\'\i}guez, Evelyn",
    title = "{Three-dimensional Maxwellian Carroll gravity theory and the cosmological constant}",
    eprint = "2107.05716",
    archivePrefix = "arXiv",
    primaryClass = "hep-th",
    doi = "10.1016/j.physletb.2021.136735",
    journal = "Phys. Lett. B",
    volume = "823",
    pages = "136735",
    year = "2021"
}

@article{Concha:2024tcu,
    author = "Concha, Patrick and Rodr{\'\i}guez, Evelyn and Salgado, Sebasti{\'a}n",
    title = "{3D Carrollian gravity from 2D Euclidean symmetry}",
    eprint = "2501.00205",
    archivePrefix = "arXiv",
    primaryClass = "hep-th",
    doi = "10.1140/epjc/s10052-025-14234-6",
    journal = "Eur. Phys. J. C",
    volume = "85",
    number = "5",
    pages = "517",
    year = "2025"
}

@article{Bergshoeff:2015ija,
    author = "Bergshoeff, Eric and Rosseel, Jan and Zojer, Thomas",
    title = "{Newton-Cartan supergravity with torsion and Schrödinger supergravity}",
    eprint = "1509.04527",
    archivePrefix = "arXiv",
    primaryClass = "hep-th",
    reportNumber = "UG-15-59, TUW-15-16, TUW-15-17",
    doi = "10.1007/JHEP11(2015)180",
    journal = "JHEP",
    volume = "11",
    pages = "180",
    year = "2015"
}

@article{Bergshoeff:2017dqq,
    author = "Bergshoeff, Eric and Chatzistavrakidis, Athanasios and Romano, Luca and Rosseel, Jan",
    title = "{Newton-Cartan Gravity and Torsion}",
    eprint = "1708.05414",
    archivePrefix = "arXiv",
    primaryClass = "hep-th",
    doi = "10.1007/JHEP10(2017)194",
    journal = "JHEP",
    volume = "10",
    pages = "194",
    year = "2017"
}

@article{VandenBleeken:2017rij,
    author = "Van den Bleeken, Dieter",
    title = "{Torsional Newton\textendash{}Cartan gravity from the large c expansion of general relativity}",
    eprint = "1703.03459",
    archivePrefix = "arXiv",
    primaryClass = "gr-qc",
    doi = "10.1088/1361-6382/aa83d4",
    journal = "Class. Quant. Grav.",
    volume = "34",
    number = "18",
    pages = "185004",
    year = "2017"
}

@article{Concha:2022you,
    author = "Concha, Patrick and Rodr\'\i{}guez, Evelyn and Rubio, Gustavo and Ya\~nez, Paola",
    title = "{Three-dimensional Newtonian gravity with cosmological constant and torsion}",
    eprint = "2204.11763",
    archivePrefix = "arXiv",
    primaryClass = "hep-th",
    doi = "10.1140/epjc/s10052-023-11210-w",
    journal = "Eur. Phys. J. C",
    volume = "83",
    number = "1",
    pages = "47",
    year = "2023"
}

@article{Concha:2023ejs,
    author = "Concha, Patrick and Merino, Nelson and Rodr{\'\i}guez, Evelyn",
    title = "{Non-relativistic limit of the Mielke{\textendash}Baekler gravity theory}",
    eprint = "2309.00500",
    archivePrefix = "arXiv",
    primaryClass = "hep-th",
    doi = "10.1140/epjc/s10052-024-12787-6",
    journal = "Eur. Phys. J. C",
    volume = "84",
    number = "4",
    pages = "407",
    year = "2024"
}

@article{Caroca:2022byi,
    author = "Caroca, Ricardo and Pe{\~n}afiel, Diego M. and Salgado-Rebolledo, Patricio",
    title = "{Nonrelativistic spin-3 symmetries in 2+1 dimensions from expanded and extended Nappi-Witten algebras}",
    eprint = "2208.00602",
    archivePrefix = "arXiv",
    primaryClass = "hep-th",
    doi = "10.1103/PhysRevD.107.064034",
    journal = "Phys. Rev. D",
    volume = "107",
    number = "6",
    pages = "064034",
    year = "2023"
}

@article{Penafiel:2019czp,
    author = "Pe{\~n}afiel, Diego M. and Salgado-Rebolledo, Patricio",
    title = "{Non-relativistic symmetries in three space-time dimensions and the Nappi-Witten algebra}",
    eprint = "1906.02161",
    archivePrefix = "arXiv",
    primaryClass = "hep-th",
    doi = "10.1016/j.physletb.2019.135005",
    journal = "Phys. Lett. B",
    volume = "798",
    pages = "135005",
    year = "2019"
}

@article{Riegler:2014bia,
    author = "Riegler, Max",
    title = "{Flat space limit of higher-spin Cardy formula}",
    eprint = "1408.6931",
    archivePrefix = "arXiv",
    primaryClass = "hep-th",
    reportNumber = "TUW-14-12, TUW--14--12",
    doi = "10.1103/PhysRevD.91.024044",
    journal = "Phys. Rev. D",
    volume = "91",
    number = "2",
    pages = "024044",
    year = "2015"
}

@article{Geiller:2020edh,
    author = "Geiller, Marc and Goeller, Christophe and Merino, Nelson",
    title = "{Most general theory of 3d gravity: Covariant phase space, dual diffeomorphisms, and more}",
    eprint = "2011.09873",
    archivePrefix = "arXiv",
    primaryClass = "hep-th",
    doi = "10.1007/JHEP02(2021)120",
    journal = "JHEP",
    volume = "02",
    pages = "120",
    year = "2021"
}

@article{Concha:2023nou,
    author = "Concha, Patrick and Fierro, Octavio and Rodr{\'\i}guez, Evelyn",
    title = "{Hietarinta Chern{\textendash}Simons supergravity and its asymptotic structure}",
    eprint = "2312.14686",
    archivePrefix = "arXiv",
    primaryClass = "hep-th",
    doi = "10.1140/epjc/s10052-024-12468-4",
    journal = "Eur. Phys. J. C",
    volume = "84",
    number = "1",
    pages = "102",
    year = "2024"
}

@phdthesis{Safari:2020pje,
    author = "Safari, H. R.",
    title = "{Deformation of Asymptotic Symmetry Algebras and Their Physical Realizations}",
    eprint = "2011.02318",
    archivePrefix = "arXiv",
    primaryClass = "hep-th",
    school = "IPM, Tehran",
    month = "9",
    year = "2020"
}

\end{document}